\documentclass[journal]{IEEEtran}
\newcommand{\ee}{\epsilon}
\usepackage{enumerate}

\usepackage{epsf,psfrag,amssymb,amsfonts,color,cite,fancybox}
\usepackage[mathscr]{eucal}
\usepackage[dvips]{graphicx}
\usepackage{ifpdf}
\usepackage{longtable}
\usepackage{multicol}
\usepackage{float}
\usepackage{bm}

\usepackage{epstopdf}

\usepackage[cmex10]{amsmath}
\usepackage[caption=false,font=footnotesize]{subfig}

\begin{document}
\renewcommand{\thefigure}{\arabic{figure}}
\renewcommand{\thesubfigure}{\alph{subfigure}}

\title{A Framework for Dynamic Stability Analysis of Power Systems with Volatile Wind Power}

\author{Xiaozhe Wang,~\IEEEmembership{Member,~IEEE,}
        Tao Wang,~\IEEEmembership{Member, ~IEEE,}
        Hsiao-Dong Chiang,~\IEEEmembership{Fellow,~IEEE,}
        Jianhui Wang,~\IEEEmembership{Senior Member,~IEEE,}
        and Hui Liu,~\IEEEmembership{Member, ~IEEE.}
    %
    %\thanks{X. Wang and J. Wang's work is partially sponsored by the U.S. Department of Energy Office of Electricity Delivery and Energy Reliability.}
    \thanks{Xiaozhe Wang is with the Department of Electrical
    and Computer Engineering, McGill University, Montreal, QC H3A 0G4, Canada. Email: xiaozhe.wang2@mcgill.ca.}
    \thanks{Tao Wang and Hsiao-Dong Chiang are with the School of
    Electrical and Computer Engineering, Cornell University, Ithaca,
    NY 14853, USA. Email: tw355@cornell.edu, hc63@cornell.edu.}
    \thanks{Jianhui Wang is with the Center for Energy, Environmental,
    and Economic Systems Analysis, Argonne National Laboratory, Argonne,
    IL 60439, USA. Email: jianhui.wang@anl.gov.}
    \thanks{Hui Liu is with the School of Electrical and
    Information Engineering, Jiangsu University, Zhenjiang, Jiangsu, China.
    Email: hughlh@126.com.}
}

\maketitle

\begin{abstract}
We propose a framework employing stochastic differential equations
to facilitate the long-term stability analysis of power grids with
intermittent wind power generations. This framework takes into
account the discrete dynamics which play a critical role in the
long-term stability analysis, incorporates the model of wind speed
with different probability distributions, and also
{develops an approximation methodology (by a
deterministic hybrid model) for the stochastic hybrid model to
reduce the computational burden brought about by the uncertainty of
wind power. The theoretical and numerical studies show
that a deterministic hybrid model can provide an accurate trajectory
approximation and stability assessments for the stochastic hybrid
model under mild conditions. In addition, we discuss the critical
cases that the deterministic hybrid model fails and discover that
these cases are caused by a violation of the proposed sufficient
conditions.} Such discussion complements the proposed framework and
methodology and also reaffirms the importance of the stochastic
hybrid model when the system operates close to its stability limit.
\end{abstract}


\IEEEpeerreviewmaketitle

\section{introduction}

Nowadays, many efforts have been devoted to producing the electric
power from renewable energy sources among which the wind power is
the most technically favorable and economically attractive \cite{Harbour:2009}.
However, volatile and uncontrollable characteristics of the wind power
generation lead to stability concerns for the secure and economic
operation of modern smart grids. As the wind penetration grows
continuously, it is imperative to investigate the impacts of wind
power generations on the system stability.

In the literature, the impacts of the wind power generation have
been studied concerning different types of stabilities
\cite{Rodriguez:2003}-\cite{Bezerra:2012}. Specifically,
\cite{Rodriguez:2003}-\cite{Loparo:2011} investigated the impacts of
different parameters (e.g., the reactive power compensation,
distance to the fault, and rotor inertia) on the transient and
frequency stabilities of a power system; \cite{Kling_pes:2002}
addressed the influence of different wind generators on the
transient stability; \cite{Harbour:2009} and \cite{Keane:2012}
studied the detrimental and beneficial influences of wind generators
on transient and small-signal stabilities by converting wind
generators to conventional synchronous generators;
\cite{Vieira:2015}\cite{Bezerra:2012} analyzed the impacts of
various control algorithms of wind generators on the long-term
stability. In those studies, the variable nature of wind power is
not considered and the wind speed is oversimplified as constant. To
address this concern, \cite{crow:2012}-\cite{Nwankpa:2000} adopted
an approach that describes the uncertainty of the wind power by
stochastic differential equations (SDEs) and investigated the
impacts of the wind generation on rotor-angle and small-signal
stabilities, in which, however, the wind power was simply modeled as
a Gaussian white noise perturbation on the power injection.

Regarding the long-term stability analysis that focuses on the time
scale when fast dynamics damp out and control devices start working,
however, a comprehensive framework is still missing in the literature to
characterize the wind power with various stochastic properties, lay
down a theoretical foundation for the stability assessment of these
stochastic systems, and develop efficient numerical tools for such
stability analysis. To address these issues, the Weibull model of
the wind speed has been incorporated into the dynamic model of the
power system to perform the long-term stability analysis
\cite{Wangxz:sde1}, {where} SDEs are applied to
describe the dynamics of the wind speed. By this SDE-based model, a
theoretical approach that approximates the stochastic model by a
deterministic model has been developed to reduce the computational
burden caused by an accurate quantification of the uncertainty. Nevertheless,
the proposed model and methodology are only applicable to continuous
power system models. On the other hand, the discrete events induced
by control and protective devices occur frequently in a long time
scale after contingencies \cite{Cutsem:book}. For instance, load tap
changers are to restore the load-side voltages; shunt compensation
switchings act to increase the transmission capability; and
OvereXcitation Limiters may be activated to protect the generators
from overheating. These discrete dynamics are generally designed to
act after the fast dynamics damp out so as to avoid unnecessary interactions with the fast dynamics
\cite{Cutsem:book}-\cite{Kundur:book}, {and they require accurate representation by discrete models in time-domain simulation \cite{Milano:2011}}. As a result, it is imperative to
integrate discrete models to perform the comprehensive
long-term stability analysis for realistic power systems.

The paper begins by showing that a power grid integrating wind
power generations can be modeled as a stochastic hybrid model (SHM),
with discrete dynamics, in a SDE-based framework in which the wind
speed model that captures various stochastic properties can be
integrated. In particular, it is analytically shown in this
framework that SHM can be approximated by a deterministic hybrid
model (DHM) which offers an accurate trajectory approximation (for
SHM) and stability assessments with high computational efficiency
if some mild sufficient conditions are satisfied. A numerical example
is presented to demonstrate the accuracy and efficiency of DHM. It
is noteworthy that SHM must be implemented whenever
{any proposed sufficient conditions are violated}.
To show this necessity, we present several numerical examples in
which DHM fails to capture the instabilities of SHM. The causes for
the failure are investigated and shown to correspond to a violation
{of} the sufficient conditions. This discussion
complements the proposed SDE-based framework, shows the application
scope of the approximation methodology, and also emphasizes the
largely-neglected necessity of the stochastic model in the long-term
stability analysis.

As the modern smart grids endeavor to incorporate high penetration of
intermittent renewable energy, integrate plug-in vehicles, and
encourage opportunistic users, the operation and control of
power grids are required to account for the resulting high
variability and uncertainty. We believe that the proposed SDE-based
framework and approximation methodology can be readily generalized
to conduct stability assessments for power systems with the
uncertainties brought about by various renewable energy sources,
plug-in vehicles, smart appliances, opportunistic users, and so
forth.

The remainder of the paper is organized as follows. Section
\ref{sectionmodel} introduces the SDE-based framework of power
system models integrating the stochastic dynamics of the wind speed.
Section \ref{sectiontheory} develops an approximation methodology
for SHM in the SDE-based framework, which provides an accurate
trajectory approximation and correct stability assessments with a
high simulation speed. In particular, a diagram is summarized at the
end of Section \ref{subsectiontheory} to illustrate the
relationships among the proposed models and theoretical results.
Furthermore, Section \ref{sectionnecessity} presents some critical
cases in which some sufficient conditions of the proposed
methodology are violated, to explain the necessity of implementing
SHM to obtain correct stability assessments.

%\section{preliminaries}\label{preliminaries}
\section{sde-based framework of hybrid models}\label{sectionmodel}

The conventional long-term stability model (i.e., the complete
dynamic model) without stochasticity for simulating the system
dynamic response to a disturbance in the $\tau$ time scale can be
described as follows {(see (22)-(25)
\cite{Wangxz:sde1} and (15) \cite{Wangxz:CAS}):}
\begin{eqnarray}
\bm{{z}_d}(k)&=&\bm{{h}_d}(\bm{z_c,x,y,z_d}(k-1))\label{slowdde}\\
\bm{{z}_{c}^\prime}&=&\bm{{h}_c({z_c,x,y,z_d})}\label{slowode}\\
\ee\bm{{{x}^\prime}}&=&\bm{f({z_c,x,y,z_d})}\label{fastode}\\
\bm{0}&=&\bm{g({z_c,x,y,z_d})}\label{algebraiceqn}
\end{eqnarray}
where $\tau=t\ee$ and $\prime$ refers to $\frac{d}{d\tau}$. Here,
(\ref{slowdde}) accounts for the long-term discrete events, such as
shunt capacitors and load tap changers (LTCs); (\ref{slowode})
depicts the slow dynamics, including self restorative loads, turbine
governors (TGs), and OvereXcitation Limiters (OXLs); (\ref{fastode})
describes the fast dynamics of components, such as synchronous
machines, doubly-fed induction generators (DFIGs), induction motors,
and exciters; and (\ref{algebraiceqn}) describes the power flow
relation and internal relationships between variables. In addition,
$\bm{h_d}$ are discrete functions; $\bm{z_d}$ are slow discrete
variables whose changing from $\bm{z_d}(k-1)$ to $\bm{z_d}(k)$
relies on (\ref{slowdde}) and occurs at times $t_k$, $1\le k\le N$.
The functions $\bm{{h}_c}$, $\bm{{f}}$, and $\bm{{g}}$ are
continuous; $\bm{z_c}$, $\bm{{x}}$, and $\bm{{y}}$ are the vectors
of slow state variables, fast state variables, and algebraic
variables, respectively; and $\ee$ is deemed as the reciprocal of
the maximum time constant among all components.

\subsection{Stochastic Model of Wind Speed}\label{wind model}

The impacts of the wind power on the system stability have been
addressed \cite{Rodriguez:2003}-\cite{Bezerra:2012} in which the
wind speed is termed as a constant and an entry of the vector
$\bm{y}$---algebraic variables. In this paper, we characterize the
randomness of the wind speed by a stochastic model.

Specifically, given $n_w$ independent wind energy sources that each
energy source follows a certain probability distribution, the wind
speeds of the $n_w$ sources are collectively denoted by a vector
$\bm{y_w}$ in the following model (see \cite{Wangxz:sde1} and
\cite{Milano:2013_1}):
\begin{eqnarray}
\ee\bm{\eta_w}^\prime &=& -A\bm{\eta_w}+\sigma\bm{\xi}={\bm{f_w}(\bm{\eta_w})+\sigma\bm{{\xi}}}, \label{wind1}\\
\bm{y_w} &=& \bm{\hat{F}_w^{-1}}\left(\bm{\hat{\Phi}}
\big(\tfrac{\bm{\eta_w}}{\sigma/\sqrt{2\bm{\alpha}}}\big)\right) =
\bm{g_w(\eta_w)}, \label{wind2}
\end{eqnarray}
where $\bm{\eta_w}, \bm{y_w} \in \mathbb{R}^{n_w}$, the matrix
$$A=\mbox{diag}(\bm{\alpha})=\mbox{diag}[\alpha_1,\ldots,\alpha_{n_w}]\in
\mathbb{R}^{{n_w}\times{n_w}}$$ determines the autocorrelation
property of $\bm{y_w}$ (see below for more details), and
$\int_0^{t}\bm{\xi}(s)ds$ is an $n_w$-dimensional Wiener process. In
addition, $\bm{\hat{F}_w}=[F_1(\eta_{w_1}),
F_2(\eta_{w_2}),...F_{n_w}(\eta_{w_{n_w}})]^T$,
$\bm{\hat{\Phi}}=[\Phi(\eta_{w_1}),\Phi(\eta_{w_2}),...\Phi(\eta_{w_{n_w}})]^T$,
and $\bm{g_w}:\mathbb{R}^{n_w}\mapsto\mathbb{R}^{n_w}$, where $F_i$
is the cumulative distribution function of the corresponding wind
speed $y_{w_i}$, and $\Phi$ is the cumulative distribution function
of a Gaussian distribution.

In model (\ref{wind1})-(\ref{wind2}) of the wind speed,
$\bm{\eta_w}$ is a vector Ornstein-Uhlenbeck process, and each
$y_{w_i}$ matches the distribution of $F_i$ by the property of the
memoryless transformation \cite{Milano:2013_1}. For example, if the
wind speed of source $n_{w_i}$ is governed by the Weibull
distribution with a shape parameter $k_i>0$ and a scale parameter
$\lambda_i>0$, then
\begin{equation}
F_{w_i}(u)={{1-e^{(u/\lambda_i)^{k_i}}}} ~\hbox{for all}~ u>0,
\end{equation}
and $y_{w_i}$ has the following statistical properties
(see (26)-(28)\cite{Milano:2013_1}):
\begin{enumerate}[(i)]
\item $\mathrm{E}[{y_{w_i}}(t)]={\lambda_i}\Gamma(1+\frac{1}{{k_i}})={\mu_{w_i}}$.

\item $\mathrm{Var}[{y_{w_i}}(t)]={\lambda_i}^2
\Gamma(1+\frac{2}{{k_i}})-\mu_{w_i}^2$.

\item $\mathrm{Aut}[{y_{w_i}}(t_k),{y_{w_i}}(t_j)]\approx
e^{-\alpha_i|t_j-t_k|}$.
\end{enumerate}
Note that $\lambda_i$, $k_i$, and $\alpha_i$ are the parameters that
determine the statistical properties of wind speed $y_{w_i}$, but
$\sigma$ does not. So $\sigma$ can be arbitrarily selected
\cite{Wangxz:sde1}\cite{Milano:2013_1}. Indeed, $\sigma$ is only
an intermediate parameter to generate the Ornstein-Uhlenbeck process
$\bm{\eta_w}$. The readers are referred to \cite{Milano:2013_1} for
more details.

\subsection{Hybrid Models}

When integrating the stochastic model (\ref{wind1})-(\ref{wind2}) of
the wind speed into the long-term stability model
(\ref{slowdde})-(\ref{algebraiceqn}), the stochastic hybrid model
(SHM) takes the following form:
\begin{eqnarray}
\bm{{z}_d}(k)&=&\bm{\bar{h}_d(\bm{z_c},\bm{\bar{x}},\bm{\bar{y}},}\bm{z_d}(k-1)\bm{)}, \label{dde}\\
\bm{{z}_{c}}^\prime&=&\bm{\bar{h}_c({\bm{z_c},\bm{\bar{x}},\bm{\bar{y}},\bm{z_d}})}, \label{slow_ode}\\
\ee{\bm{\bar{x}}}^\prime&=&\bm{\bar{f}({\bm{z_c},\bm{\bar{x}},\bm{\bar{y}},\bm{z_d}})}+\sigma B \bm{\bar{\xi}}, \label{fast_ode}\\
\bm{0}&=&\bm{\bar{g}({\bm{z_c},\bm{\bar{x}},\bm{\bar{y}},\bm{z_d}})},
\label{alg eqn}
\end{eqnarray}
{where $\bm{\bar{x}} \doteq
\left[\begin{smallmatrix}\bm{x}
\\ \bm{\eta_w}\end{smallmatrix}\right]$,
$\bm{\bar{y}} \doteq \left[\begin{smallmatrix}\bm{y}
\\ \bm{y_w}\end{smallmatrix}\right]$, and
$B \doteq
\left[\begin{smallmatrix}0\\I_{n_w}\end{smallmatrix}\right]$,
nonzero entries of which correspond to $n_w$ independent wind
sources. In addition, $\bm{\bar{f}}\doteq \left[\begin{smallmatrix} \bm{f} \\
\bm{f_w}
\end{smallmatrix}\right]$, $\bm{\bar{g}} \doteq  \left[\begin{smallmatrix} \bm{g} \\
\bm{p}
\end{smallmatrix}\right]$ with $\bm{p} \doteq \bm{y_w}-\bm{g_w(\eta_w)}$,
and $\bar{\bm{\xi}}=\left[\begin{smallmatrix}
0\\ {\bm{\xi}}\end{smallmatrix}\right]\in \mathbb{R}^{{n_x}\times{n_w}}$.
%where $\bar{h}_c:
%\mathbb{R}^{n_{z_c}}\times\mathbb{R}^{n_{x}}\times\mathbb{R}^{n_{\bar{y}}}\times\mathbb{R}^{n_{{z_d}}}\mapsto
%\mathbb{R}^{n_{z_c}}$, $\bar{f}:
%\mathbb{R}^{n_{z_c}}\times\mathbb{R}^{n_{\bar{x}}}\times\mathbb{R}^{n_{\bar{y}}}\times\mathbb{R}^{n_{{z_d}}}
%\mapsto \mathbb{R}^{n_{\bar{x}}}$, and
%$\bar{g}:\mathbb{R}^{n_{z_c}}\times\mathbb{R}^{n_{\bar{x}}}\times\mathbb{R}^{n_{\bar{y}}}\times\mathbb{R}^{n_{{z_d}}}
%\mapsto \mathbb{R}^{n_{\bar{y}}}$.
Here, (\ref{dde}) and
(\ref{slow_ode}) are directly derived from (\ref{slowdde}) and
(\ref{slowode}), respectively, such that
$\bm{\bar{h}_d(\bm{z_c},\bm{\bar{x}},\bm{\bar{y}},}\bm{z_d}(k-1)\bm{)}
= \bm{{h}_d}(\bm{z_c,x,y,z_d}(k-1))$ and
$\bm{\bar{h}_c({\bm{z_c},\bm{\bar{x}},\bm{\bar{y}},\bm{z_d}})}=
\bm{{h}_c({z_c,x,y,z_d})}$; (\ref{fast_ode}) is obtained from a
combination of (\ref{fastode}) and (\ref{wind1}), whereas (\ref{alg
eqn}) is derived by combining (\ref{algebraiceqn}) and
(\ref{wind2}).}

%\textcolor{blue}{where (\ref{dde}) and (\ref{slow_ode}) are
%identical to (\ref{slowdde}) and (\ref{slowode}), respectively,
%(\ref{fast_ode}) is obtained from a combination of (\ref{fastode})
%and (\ref{wind1}), whereas (\ref{alg eqn}) is derived by combining
%(\ref{algebraiceqn}) and (\ref{wind2}). Here,
%$\bm{\bar{x}}=[\bm{x}^T,\bm{\eta_w}^T]^T$,
%$\bm{\bar{y}}=[\bm{y}^T,\bm{y_w}^T]^T$, and
%$B=\left[\begin{smallmatrix}0\\I_{n_w}\end{smallmatrix}\right]$, the
%nonzero entries of which correspond to $n_w$ independent wind
%sources. In addition, $\bm{\bar{g}} = \left[\begin{smallmatrix} \bm{g} \\
%\bm{p}
%\end{smallmatrix}\right]$ with $\bm{p}=\bm{y_w}-\bm{g_w(\eta_w)}$,
%and $\bar{\xi}=\left[\begin{smallmatrix}
%0\\\xi\end{smallmatrix}\right]\in \mathbb{R}^{{n_x}\times{n_w}}$,
%where $\bar{h}_c:
%\mathbb{R}^{n_{z_c}}\times\mathbb{R}^{n_{x}}\times\mathbb{R}^{n_{\bar{y}}}\times\mathbb{R}^{n_{{z_d}}}\mapsto
%\mathbb{R}^{n_{z_c}}$, $\bar{f}:
%\mathbb{R}^{n_{z_c}}\times\mathbb{R}^{n_{\bar{x}}}\times\mathbb{R}^{n_{\bar{y}}}\times\mathbb{R}^{n_{{z_d}}}
%\mapsto \mathbb{R}^{n_{\bar{x}}}$, and
%$\bar{g}:\mathbb{R}^{n_{z_c}}\times\mathbb{R}^{n_{\bar{x}}}\times\mathbb{R}^{n_{\bar{y}}}\times\mathbb{R}^{n_{{z_d}}}
%\mapsto \mathbb{R}^{n_{\bar{y}}}$.}

Recall that discrete dynamics described by (\ref{dde}) play
important roles in the long-term stability because many protective
and control devices may take effect in the long-term time scale to
restore the load-sided power, protect generators, and so on.

This study aims to show that the SHM (\ref{dde})-(\ref{alg eqn}) can
be well approximated by a deterministic hybrid model (DHM), say,
\begin{eqnarray}
\bm{{z}_d}(k) &=& \bm{\bar{h}_d(\bm{z_c},\bm{\bar{x}},\bm{\bar{y}},}\bm{z_d}(k-1)\bm{)}, \label{det_dde}\\
\bm{{z}_{c}}^\prime &=& \bm{\bar{h}_c({\bm{z_c},\bm{\bar{x}},\bm{\bar{y}},\bm{z_d}})}, \label{det_slow_ode}\\
\ee{\bm{\bar{x}}}^\prime &=& \bm{\bar{f}({\bm{z_c},\bm{\bar{x}},\bm{\bar{y}},\bm{z_d}})}, \label{det_fast_ode}\\
\bm{0} &=&
\bm{\bar{g}({\bm{z_c},\bm{\bar{x}},\bm{\bar{y}},\bm{z_d}})}.
\label{det_alg eqn}
\end{eqnarray}

Note that the vector of algebraic variables $\bm{\bar{y}}$ in
(\ref{alg eqn}) and (\ref{det_alg eqn}) can be eliminated under
\textit{Assumption 1} {which is a generic property
satisfied in normal operating conditions
\cite{Cutsem:book}\cite{Chiang:book}.} \\

\textbf{\textit{Assumption 1.}} {\it The DHM
(\ref{det_dde})-(\ref{det_alg eqn}) does not encounter singularity,
i.e., $\frac{\partial{\bm{\bar{g}}}}{\partial{\bm{y}}}$ is
nonsingular along the trajectory.}\\

%Assuming $\partial_{\bar{y}}\bar{g}$ is nonsingular, then according to implicit function theorem, Substituting the expression back to model (\ref{dde})-(\ref{alg eqn}),
Under \textit{Assumption 1}, $\bm{\bar{y}}$ can be represented in
terms of $\bm{z_c}$, $\bm{\bar{x}}$, and $\bm{z_d}$ using
(\ref{det_alg eqn}), namely
$\bm{\bar{y}=m(\bm{z_c},\bm{\bar{x}},\bm{z_d})}$. Then, the SHM
(\ref{dde})-(\ref{alg eqn}) can be written as:
\begin{eqnarray}
\bm{z_d}(k)&=&\bm{H_d(\bm{z_c},\bm{\bar{x}},}\bm{z_d}(k-1)\bm{)}, \label{sde power dde}\\%\label{sde power dde-0}\\
\bm{z_c}^\prime&=&\bm{H_c(\bm{z_c},\bm{\bar{x}},\bm{z_d})}, \label{sde power slow}\\%\label{sde power slow-0}\\
\ee\bm{\bar{x}}^\prime&=&\bm{F(\bm{z_c},\bm{\bar{x}},\bm{z_d})}+\sigma B {\bm{\bar{\xi}}}. \label{sde power fast}%\label{sde power fast-0}
\end{eqnarray}
By analogy, the DHM (\ref{det_dde})-(\ref{det_alg eqn}) is
equivalently converted to:
\begin{eqnarray}
\bm{z_d}(k)&=&\bm{H_d(\bm{z_c},\bm{\bar{x}},}\bm{z_d}(k-1)\bm{)}, \label{det power dde}\\
\bm{z_c}^\prime &=&\bm{H_c(\bm{z_c},\bm{\bar{x}},\bm{z_d})}, \label{det power slow}\\
\ee\bm{\bar{x}}^\prime &=&\bm{F(\bm{z_c},\bm{\bar{x}},\bm{z_d})}.
\label{det power fast}
\end{eqnarray}

In section \ref{sectiontheory}, a theoretical foundation is to be
developed to ensure the effectiveness of the approach that
approximates the SHM (\ref{sde power dde})-(\ref{sde power fast}) by
the DHM (\ref{det power dde})-(\ref{det power fast}) in the
long-term stability study. The key is to show that if some mild
conditions are satisfied, then the DHM (\ref{det power
dde})-(\ref{det power fast}) is theoretically ensured to provide an
accurate trajectory approximation and stability assessments for the
SHM (\ref{sde power dde})-(\ref{sde power fast}). Clearly, the DHM
consumes much less computational resources in the simulation
compared with the SHM and may serve as an efficient stability
assessment tool for power grids with significant wind power
generations.

\section{an approximation methodology for stochastic hybrid model}\label{sectiontheory}

The singular perturbation method for SDEs
\cite{Gentz:2006}-\cite{Freidlin:book} and sufficient conditions for
the quasi steady-state (QSS) model
\cite{Wangxz:CAS}\cite{Wangxz:wiley} are employed here to develop a
theoretical foundation for an approximation of the SHM (\ref{sde
power dde})-(\ref{sde power fast}) by the DHM (\ref{det power
dde})-(\ref{det power fast}). A numerical example using a 145-bus system is presented to demonstrate the
accuracy and efficiency of the DHM.

\subsection{Theoretical Foundation}\label{subsectiontheory}

In the SHM, when the discrete jumping is initiated, discrete
variables $\bm{z_d}$ are updated first by (\ref{sde power dde}), and
then the system acts according to (\ref{sde power slow})-(\ref{sde
power fast}) with constant $\bm{z_d}$. In this regard, one can treat
the SHM (\ref{sde power dde})-(\ref{sde power fast}) as a series of
continuous systems (\ref{sde power slow})-(\ref{sde power fast})
with constant $\bm{z_d}$ \cite{Cutsem:book}. Similarly, the DHM
(\ref{det power dde})-(\ref{det power fast}) can be considered as a
series of continuous systems (\ref{det power slow})-(\ref{det power
fast}) with constant $\bm{z_d}$. It is reasonable to assume that the
SHM and the DHM are governed by the same sequence of parameter
values $\bm{z_d}$ given the same initial condition. So, the hybrid
models (i.e., SHM and DHM) can be analyzed by comparing the
corresponding continuous systems in the series. Additionally, we
suppose that each deterministic continuous system (\ref{det power
slow})-(\ref{det power fast}) satisfies some generic
differentiability and non-degeneracy conditions (see
\textit{Assumption 2.1} \cite{Wangxz:sde1}), which are reasonable
assumptions for real-life physical systems.

\begin{figure}[!ht]
\centering
\includegraphics[width=2.5in,keepaspectratio=true,angle=0]{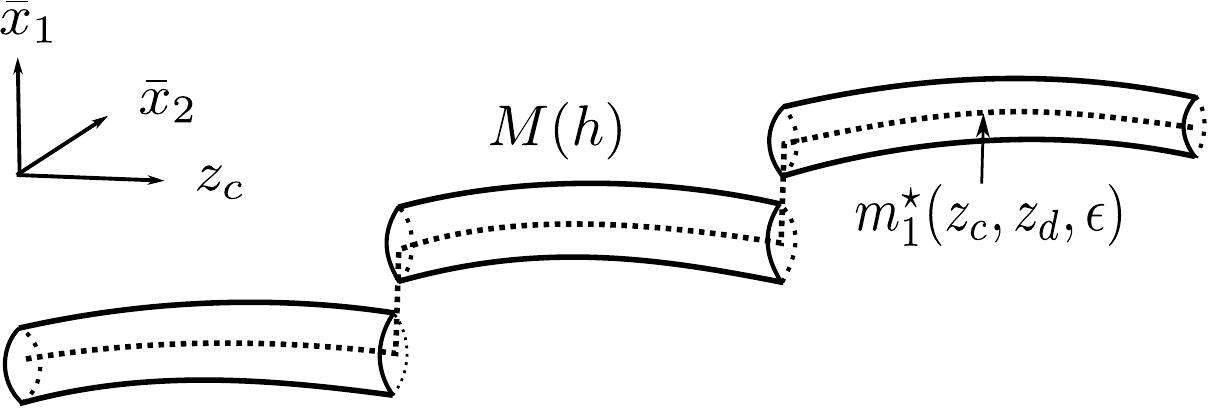}
\caption{An illustrtion of $M(h)$ in the DHM. Here, $n_{z_c}=1$,
$n_{\bar{x}}=2$, and $M(h)$ is an ellipsoidal layer around
$\bm{m_1^\star}(\bm{z_c},\bm{z_d},\ee)$. }\label{N(h)}
\end{figure}

If $\bm{\bar{x}}={\bm{m_1(\bm{z_c},\bm{z_d})}}$ is an asymptotically
stable
 equilibrium point of the short-term stability model
 $\bm{0}=\bm{F(\bm{z_c},\bm{\bar{x}},\bm{z_d})}$ for all $\bm{z_c}$ and $\bm{z_d}$,
i.e., $\bm{\bar{x}}=\bm{m_1(\bm{z_c},\bm{z_d})}$ is a stable
component of the constraint manifold, then there exists an invariant
manifold of system (\ref{det power dde})-(\ref{det power fast}):
$\bm{\bar{x}}=\bm{m_1^\star}(\bm{z_c},\bm{z_d},\ee)=\bm{m_1(\bm{z_c},\bm{z_d})}+O(\ee)$
for sufficiently small $\ee$
\cite{Wangxz:sde1}\cite{Khalil:book}\cite{Alberto:article}, where
$\bm{m_1(\bm{z_c},\bm{z_d})}$ and
$\bm{m_1^\star}(\bm{z_c},\bm{z_d},\ee)$ can be not smooth. An
ellipsoidal layer ${M}(h)$ around
$\bm{m_1^\star}(\bm{z_c},\bm{z_d},\ee)$ is defined as follows:
\begin{eqnarray}
{M}(h)&\doteq&\{(\bm{z_c},\bm{\bar{x}},\bm{z_d}):\langle (\bm{\bar{x}}-\bm{m_1}^\star(\bm{z_c},\bm{z_d},\ee)),\nonumber\\
&&\quad
M_1^\star(\bm{z_c},\bm{z_d},\ee)^{-1}(\bm{\bar{x}}-\bm{m_1^\star}(\bm{z_c},\bm{z_d},\ee))\rangle<
h^2\}. \nonumber
\end{eqnarray}
Here, the matrix $M_1^\star(\bm{z_c},\bm{z_d},\ee)$ that represents
the cross section of $M(h)$ is properly defined (see Appendix B in
\cite{Wangxz:sde1}), and an illustration for $M(h)$ is shown in Fig.
\ref{N(h)}.

Given the well-defined initial condition for each continuous system
(\ref{sde power slow})-(\ref{sde power fast}) with constant
$\bm{z_d}$ (of the SHM), the following theorem shows that the
trajectories of the SHM (\ref{sde power dde})-(\ref{sde power fast})
are confined in $M(h)$ despite the changes of discrete variables,
provided that the slow manifold is stable.\\

\textbf{\textit{Theorem 1 (Sample-Path Concentration for SHM):}}\\
{\it Consider the SHM (\ref{sde power dde})-(\ref{sde power fast})
in the study region $D_{\bm{z_d}} \times D_{\bm{z_c}} \times
D_{\bm{x}}$, for some fixed $\ee_0>0$, $h_0>0$, there exist
$\delta_0>0$, a time $\tilde{\tau}_k$ of order $\ee|\mbox{log}h|$,
\textcolor{black}{and $\tau_k > \tilde{\tau}_k$}  for each continuous
system (\ref{sde power slow})-(\ref{sde power fast}) with $0\le k\le
N$ such that if the following conditions (i) and (ii) are satisfied:

\begin{description}
\item[(i)] The slow manifold $\bm{\bar{x}}=\bm{m_1(\bm{z_c},\bm{z_d})}$ is a
stable component of the constraint manifold, where $\bm{z_c}\in
D_{\bm{z_c}}$ and $\bm{z_d}\in D_{\bm{z_d}}$;

\item[(ii)] The initial condition $(\bm{z_c}^k(0),\bm{\bar{x}}^k(0),\bm{z_d}(k))$ for {each} continuous
system (\ref{sde power slow})-(\ref{sde power fast}) of the SHM
satisfies that $(\bm{z_c}^k(0),\bm{\bar{x}}^k(0),\bm{z_d}(k))\in
M(\delta_0)$, where $\bm{z_c}^k(0)\in D_{\bm{z_c}}$ and
$\bm{z_d}(k)\in D_{\bm{z_d}}$ for $k\in[0,1,...N]$,
\end{description}

then, for all $\tau\in
\Pi=\cup_{i=1}^{N-1}[\tilde{\tau}_i,\tau_i)\cup[\tilde{\tau}_N,\infty]$,
the sample path $(\bm{z_c}(\tau),\bm{\bar{x}}(\tau),\bm{z_d}(\tau))$
of the SHM (\ref{sde power dde})-(\ref{sde power fast}) satisfies
the following probability property:
\begin{eqnarray}\label{theorem1}
&\mathbb{P}\{\exists\, \tau\in\Pi: &(\bm{z_c}(\tau),\bm{\bar{x}}(\tau),\bm{z_d}(\tau))\notin M(h)\}\nonumber\\
&&\leq
C_{n_{z_c},n_x}(\tau,\ee)e^{\frac{-h^2}{2\sigma^2}(1-O(h)-O(\ee))},
\end{eqnarray}
for all $\ee\leq\ee_0$, $h\leq h_0$, where the coefficient
$C_{n_{z_c},n_x}(\tau,\ee)=[C^{n_{z_c}}+h^{-n_x}](1+\frac{\tau}{\ee^2})$
is linear in $\tau$.}

\vspace{12pt}

{Proof:} See Appendix \ref{prooftheorem1}.\\

\textit{Theorem 1} shows that if conditions (i)-(ii) are satisfied,
then the probability that the sample path leaves $M(h)$ is less than
the right hand side (RHS) of (\ref{theorem1}). Specifically, if
$h\gg \sigma$, i.e., the deepness of the layer $h$ is far larger
than $\sigma$ related with wind speeds, then the RHS of
(\ref{theorem1}) becomes very small, which suggests that the sample
pathes of the SHM do not leave $M(h)$ almost surely
\cite{Wangxz:sde1}\cite{Gentz:2006}. So, there is no need to worry
about the probability when investigating the relations between the
trajectory of the SHM (\ref{sde power dde})-(\ref{sde power fast})
and that of the DHM (\ref{det power dde})-(\ref{det power fast}). On
the other hand, $\sigma$ does not influence the stochastic
properties of wind speed $\bm{y_w}$ as stated in Section \ref{wind
model} or Section III-A in \cite{Wangxz:sde1} (where $\sigma$ is
only an intermediate parameter to generate the Ornstein-Uhlenbeck
process $\bm{\eta_w}$). In this regard, $\sigma$ can be selected as
small as needed such that any adequate $h$ satisfies $h\gg\sigma$.
In other words, the requirement $h\gg\sigma$ can be readily
fulfilled in this SDE-based framework. {In addition, Theorem 2.4
\cite{Gentz:2003} has commented that for $h\gg\sigma$, the first
exit time that the solution $\bm{z_c}$ of the (continuous)
stochastic system (\ref{sde power slow}) leaves the region
$D_{\bm{z_c}}$ is very large (exponentially in $h^2/\sigma^2$), that
is, $\bm{z_c}$ still stays within $D_{\bm{z_c}}$ almost for sure
right before the (discrete) change of $\bm{z_d}$ occurs at
$t_k$. Note that, for adequately controlled systems, discrete
devices generally do not result in severe perturbations to the
system dynamics. So, these facts suggest that condition (ii) in
Theorem 1 is generally satisfied under normal operating conditions.}

\begin{figure}[!ht]
\centering
\includegraphics[width=3in,keepaspectratio=true,angle=0]{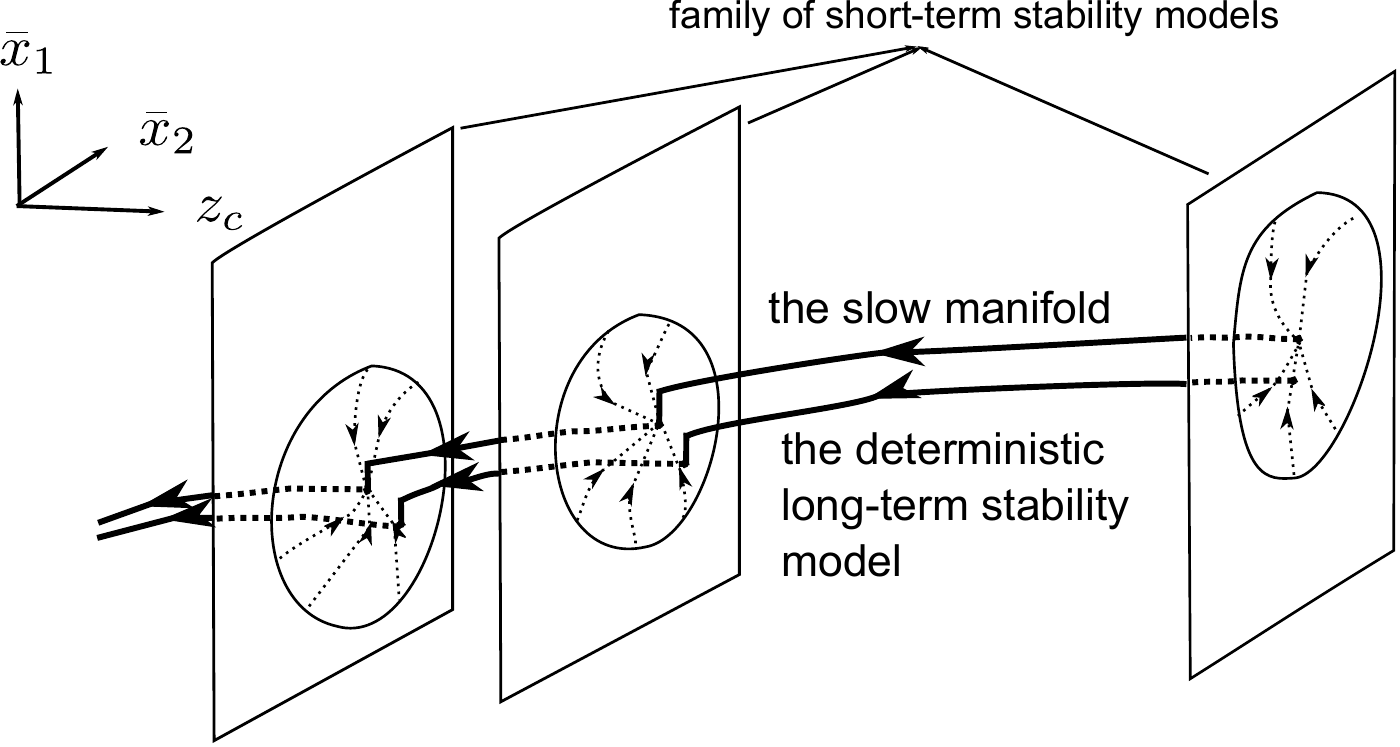}
\caption{The situation when the DHM satisfies the condition of
consistent attraction.}\label{consistentattraction}
\end{figure}

Under the condition $h\gg \sigma$, we next investigate the
relationship between the trajectory of the SHM (\ref{sde power
dde})-(\ref{sde power fast}) and that of the DHM (\ref{det power
dde})-(\ref{det power fast}). If (a) the trajectory of the SHM
remains in $M(h)$ which is an $\ee$ neighborhood of the invariant
manifold $\bm{m_1^\star}(\bm{z_c},\bm{z_d},\ee)$, and (b) the
trajectory of the DHM evolves along
$\bm{m_1^\star}(\bm{z_c},\bm{z_d},\ee)$, then we show that the
distance between the trajectory of the SHM and that of the DHM can
be readily obtained. Note that \textit{Theorem 1} provides
sufficient conditions for (a). So, the remaining question is about
how to ensure (b). Incidentally, the theoretical foundation for the
quasi steady-state (QSS) model in \cite{Wangxz:CAS} has provided
sufficient conditions for (b). In particular, one of the sufficient
conditions for (b) is the condition of \textit{consistent
attraction} defined below and illustrated in Fig.
\ref{consistentattraction}.\\

\textbf{\textit{Definition 1. Consistent Attraction
\cite{Wangxz:CAS}}}: {\it By fixing $\bm{z_c}$ and $\bm{z_d}$ as the
parameters, the short-term stability model refers to (\ref{det power
fast}). We say that the DHM (\ref{det power dde})-(\ref{det power
fast}) satisfies the condition of \textit{consistent attraction} if
the initial condition is contained in the stability region of the
initial short-term stability model and whenever discrete variables
jump from $\bm{z_d}(k-1)$ to $\bm{z_d}(k)$, $k=1,2,...,N$, the point
on trajectory of the DHM immediately after $\bm{z_d}$ jump still
stays within the stability region of the corresponding short-term
stability model.}

\textit{The condition of consistent attraction} ensures that the
trajectory of the DHM is always close to the slow manifold
$\bm{m_1(\bm{z_c,z_d})}$ despite the changing of discrete variables
(if the slow manifold is also stable), then the trajectory of the
DHM always evolves along the invariant manifold
$\bm{m_1^\star}(\bm{z_c},\bm{z_d},\ee)$. Let
$(\bm{z_{c}}(\tau),\bm{\bar{x}}(\tau),\bm{z_d}(\tau))$ be the
trajectory of the SHM (\ref{sde power dde})-(\ref{sde power fast}),
and let $(\bm{z_{cD}}(\tau),\bm{\bar{x}_D}(\tau),\bm{z_d}(\tau))$ be
that of the DHM (\ref{det power dde})-(\ref{det power fast}). Then,
the following theorem reveals the relationship between the
trajectories of the two models.\\

\textbf{\textit{Theorem 2 (Trajectory Relationship for Hybrid
Models):}}\\ {\it Given $h\gg\sigma$, consider the SHM (\ref{sde
power dde})-(\ref{sde power fast}) and the DHM (\ref{det power
dde})-(\ref{det power fast}) in the study region $D_{\bm{z_d}}
\times D_{\bm{z_c}} \times D_{\bm{x}}$, for some fixed
$\ee_0\in(0,h)$, there exist $\delta_0>0$, a time
$\tilde{\tau}_k$ of order $\ee|\mbox{log}h|$ and
$\bar{\tau}_k>\tilde{\tau}_k$ for each continuous system (\ref{sde
power slow})-(\ref{sde power fast}) where $k=0,1,...,N$, such that if
the following conditions (i), (ii) and (iii) are satisfied:
\begin{description}
\item[(i)] The slow manifold $\bm{\bar{x}}=\bm{m_1(\bm{z_c},\bm{z_d})}$ is a
stable component of the constraint manifold, where $\bm{z_c}\in
D_{\bm{z_c}}$ and $\bm{z_d}\in D_{\bm{z_d}}$;

\item[(ii)] The initial condition $(\bm{z_c}^k(0),\bm{\bar{x}}^k(0),\bm{z_d}(k))$ for each continuous
system (\ref{sde power dde})-(\ref{sde power fast}) of the SHM
satisfies $(\bm{z_c}^k(0),\bm{\bar{x}}^k(0),\bm{z_d}(k))\in
M(\delta_0)$, where $\bm{z_c}^k(0)\in D_{\bm{z_c}}$ and
$\bm{z_d}(k)\in D_{\bm{z_d}}$ for $k\in[0,1,...N]$;

\item [(iii)] The DHM (\ref{det power dde})-(\ref{det power fast})
satisfies \textbf{the condition of consistent attraction},
\end{description}
then, for $\tau\in\cup_{i=1}^{N}[\tilde{\tau}_i,\bar{\tau}_i]$, the
following relations hold:
\begin{eqnarray}
|\bm{\bar{x}}(\tau)-\bm{\bar{x}_D}(\tau)|&=&O(\sigma), \label{corollary3_1}\\
|\bm{z_{c}}(\tau)-\bm{z_{cD}}(\tau)|&=&O(\sigma\sqrt{\ee}),
\label{corollary3_2}
\end{eqnarray}
for all $\ee\in(0,\ee_0)$.}

\vspace{12pt}

Proof: See Appendix \ref{prooftheorem2}.\\

\begin{figure}[!ht]
\centering
\includegraphics[width=3in,keepaspectratio=true,angle=0]{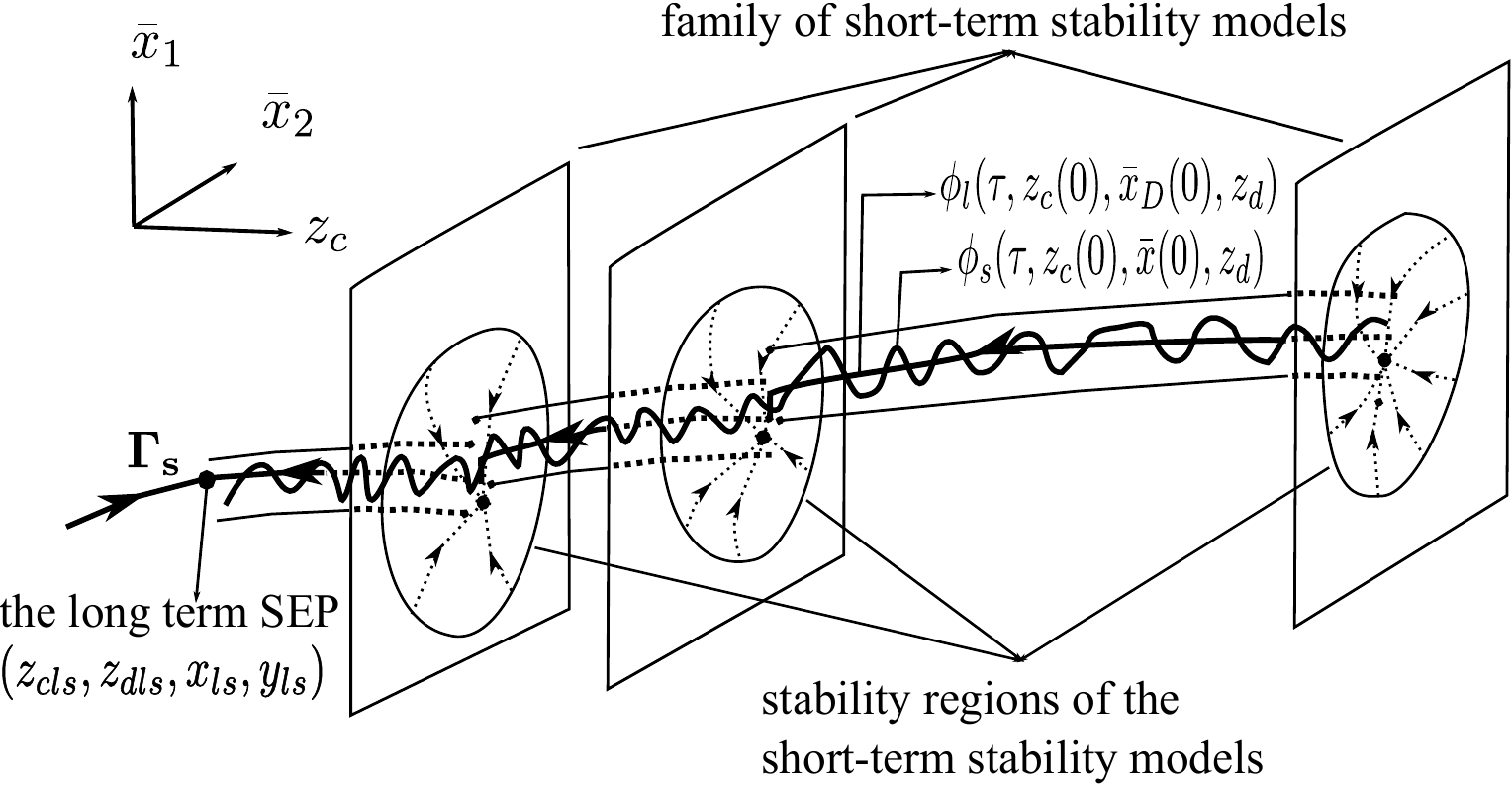}
\caption{The trajectory
$\phi_s(\tau,\bm{z_c}(0),\bm{\bar{x}}(0),\bm{z_d})$ of SHM is
bounded in $M(\sigma)$, and can be estimated by the trajectory
$\phi_l(\tau,\bm{z_c}(0),\bm{\bar{x}}_D(0),\bm{z_d})$ of DHM.
}\label{stochasticdeterministic}
\end{figure}

By \textit{Theorem 2} we observe that if the proposed sufficient
conditions are satisfied, the trajectory of the SHM can be
approximated by that of the DHM as illustrated in Fig.
\ref{stochasticdeterministic}.

\begin{figure*}[!ht]
  \includegraphics[width=7in]{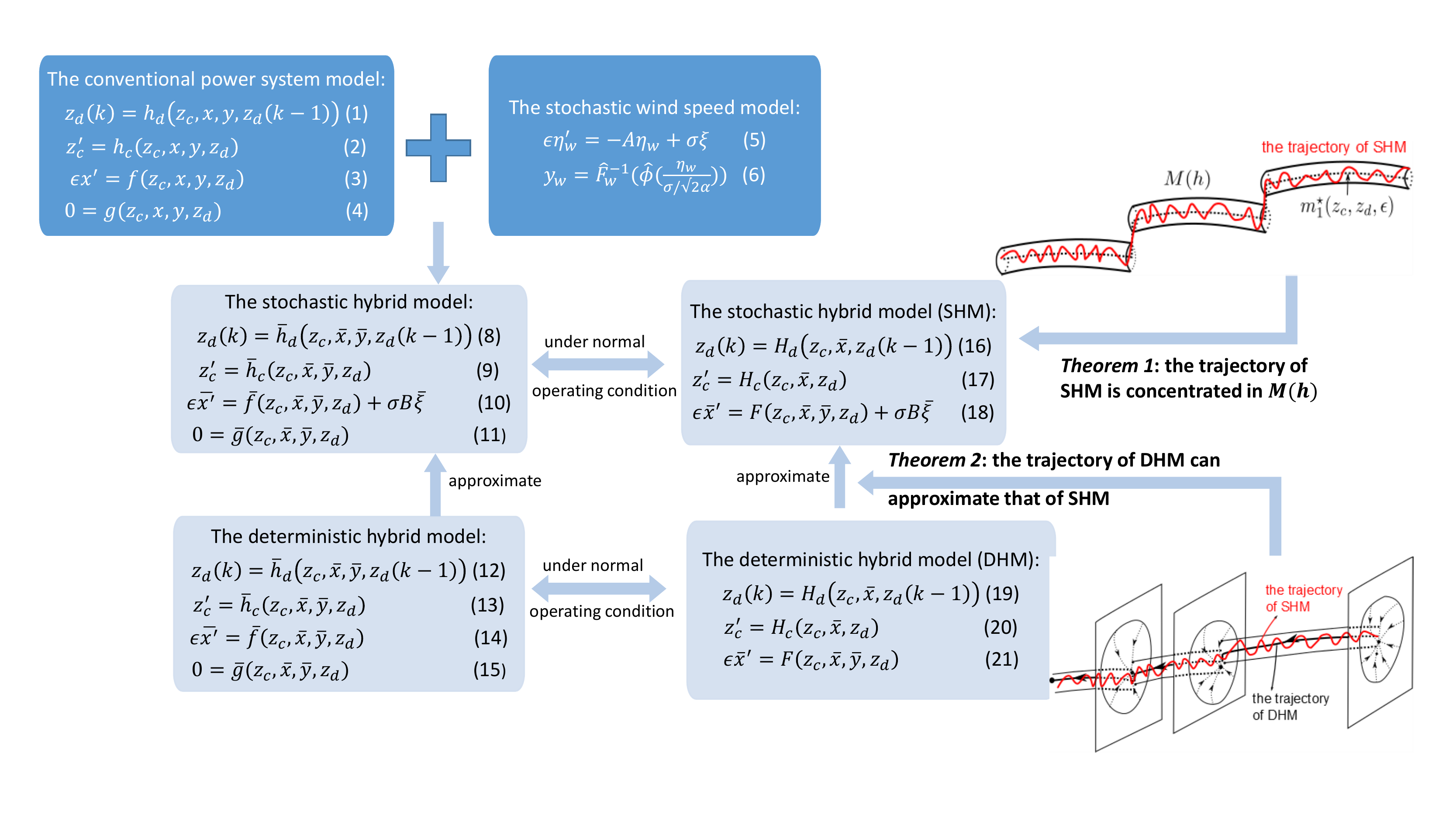}
  \caption{The proposed SDE-based framework and the
  approximation methodology.}\label{framework}
\end{figure*}

Generally speaking, sufficient conditions (i)-(iii) in
\textit{Theorem 2} are moderate and satisfied when the system
operates away from the stability boundary, and thus the DHM can
substitute the SHM and typically offer correct stability assessments
with less simulation time. But, as detailed in Section
\ref{sectionnecessity}, the SHM must be applied if any of the
sufficient conditions is violated.

For clarity, we summarize in Fig. \ref{framework} the proposed
SDE-based framework, relationship between different models, and
importance of derived theoretical results. In the SDE-based
framework, the stochastic model (\ref{wind1})-(\ref{wind2}) of the
wind speed is incorporated into the conventional power system model
(\ref{slowdde})-(\ref{algebraiceqn}), and the resulting hybrid model
(\ref{dde})-(\ref{alg eqn}) is equivalent to the SHM (\ref{sde power
dde})-(\ref{sde power fast}) under normal operating condition.
Specifically, \textit{Theorem 1-2} shows that the DHM (\ref{det
power dde})-(\ref{det power fast}) can well approximate the SHM
(\ref{sde power dde})-(\ref{sde power fast}). In particular,
\textit{Theorem 1} suggests that the sample paths of (\ref{sde power
dde})-(\ref{sde power fast}) are concentrated in a neighborhood
$M(h)$ of the invariant manifold
$\bm{m_1^\star}(\bm{z_c},\bm{z_d},\ee)$, while \textit{Theorem 2}
asserts that the DHM (\ref{det power dde})-(\ref{det power fast})
can provide an accurate trajectory approximation and stability
assessments for the SHM (\ref{sde power dde})-(\ref{sde power fast})
under some mild conditions. So, under normal operating conditions
and the proposed mild conditions, the DHM
(\ref{det_dde})-(\ref{det_alg eqn}) can well approximate the SHM
(\ref{dde})-(\ref{alg eqn}) in terms of the trajectory and stability
assessments.

\subsection{Numerical Illustration}

Numerical studies using a 145-bus test case \cite{testcasearchive}
are conducted in PSAT-2.1.8 \cite{Milano:PSAT} to show the accuracy
and efficiency of the derived results. The test system has 6
doubly-fed induction generators (DFIGs) driven by 6 independent
Weibull-distributed wind sources. The parameters of Weibull
distributions are referred from \cite{Milano:2013_1} which fit the
1-h wind speed data of the Cape St. James and Victoria Airport. The
readers are referred to Table 1 \cite{Milano:2013_1} for more
details. In addition, there are 50 synchronous generators (GENs)
with automatic voltage regulators (AVRs). Turbine governors (TGs)
are equipped for GEN 10-GEN 20, and OvereXcitation Limiters (OXLs)
are also equipped for GEN 1-GEN 6. The initial time delays of OXLs
are 50s. Moreover, 5 discrete load tap changers (LTCs) are installed
at Bus 79-95, Bus 1-33, Bus 79-92, Bus 1-5, and Bus 60-95,
respectively. Particularly, the discrete model of LTCs is shown
below \cite{Kundur:book}:
\begin{equation}\label{LTC_d1}
n(k+1)=\left\{\begin{array}{ll}n(k)+\triangle{n}, &\mbox{if  }v>v_0+d\mbox{  and  }n(k)<n^{max};\\
n(k)-\triangle{n}, &\mbox{if  }v<v_0-d\mbox{  and  }n(k)>n^{min};\\
n(k), &\mbox{otherwise};\end{array}\right.
\end{equation}
where $n$ is the tap changer ratio, $v$ is the controlled voltage,
$v_0$ is the reference voltage, $d$ is half of the LTC dead-band,
$n^{max}$ and $n^{min}$ are the upper and lower tap limits,
respectively. All LTCs have initial time delays of 50s and fixed
tapping delays of 10s. At 0.5s, three lines at Bus 95-138, Bus
94-138, Bus 94-95 trip.

{Note that the dynamic
models for synchronous generators and DFIGs used in this and
subsequent numerical examples are all detailed in Ch.\,17 and
Ch.\,21 \cite{Milano:PSATDOC}. Specifically, the order II and order IV
models of GENs are employed for the simulation of this 145-bus system.}

\begin{figure}[!ht]
\centering
\begin{minipage}[t]{0.5\linewidth}
\includegraphics[width=1.8in ,keepaspectratio=true,angle=0]{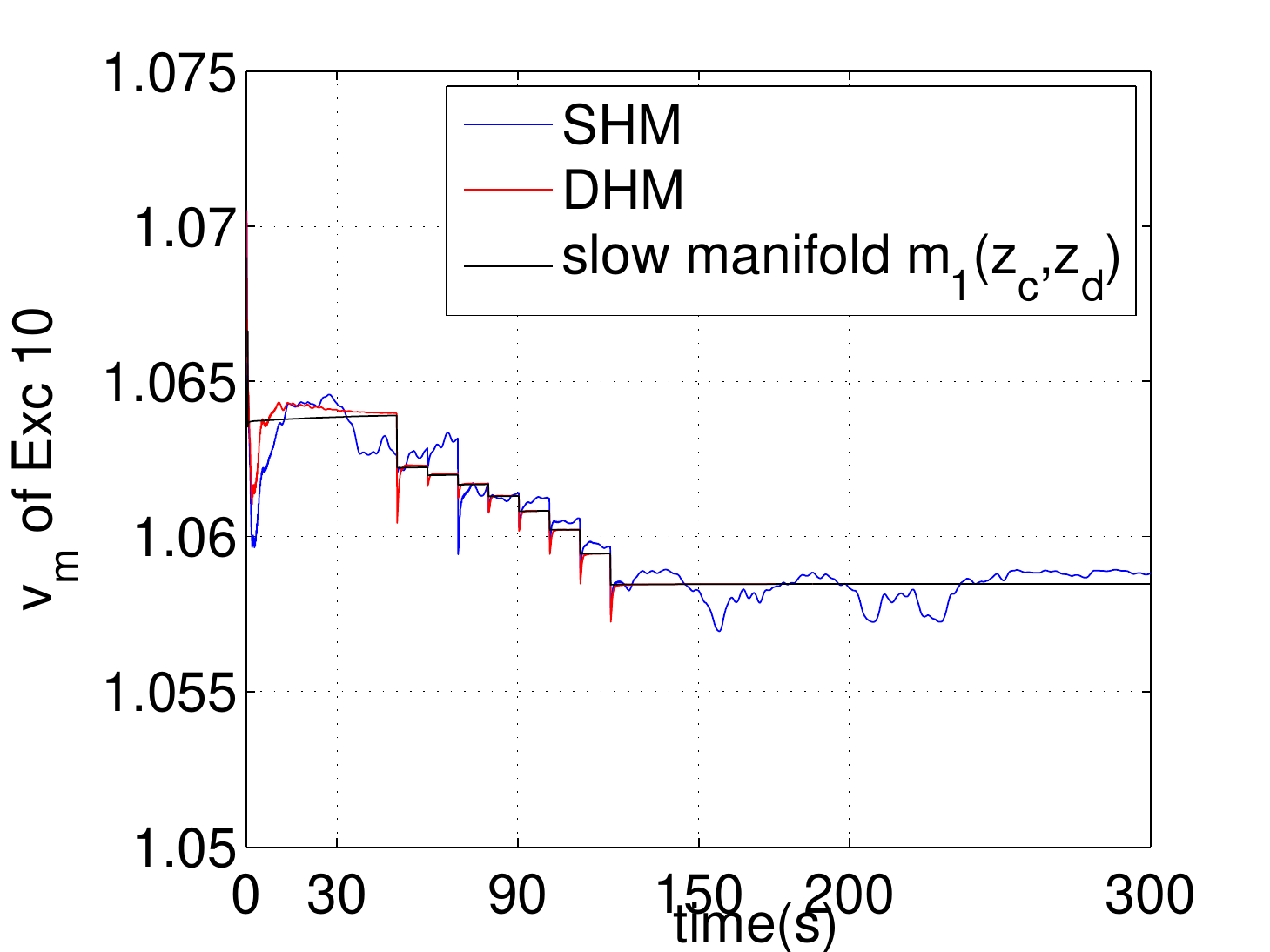}
\end{minipage}%
\begin{minipage}[t]{0.5\linewidth}
\includegraphics[width=1.8in ,keepaspectratio=true,angle=0]{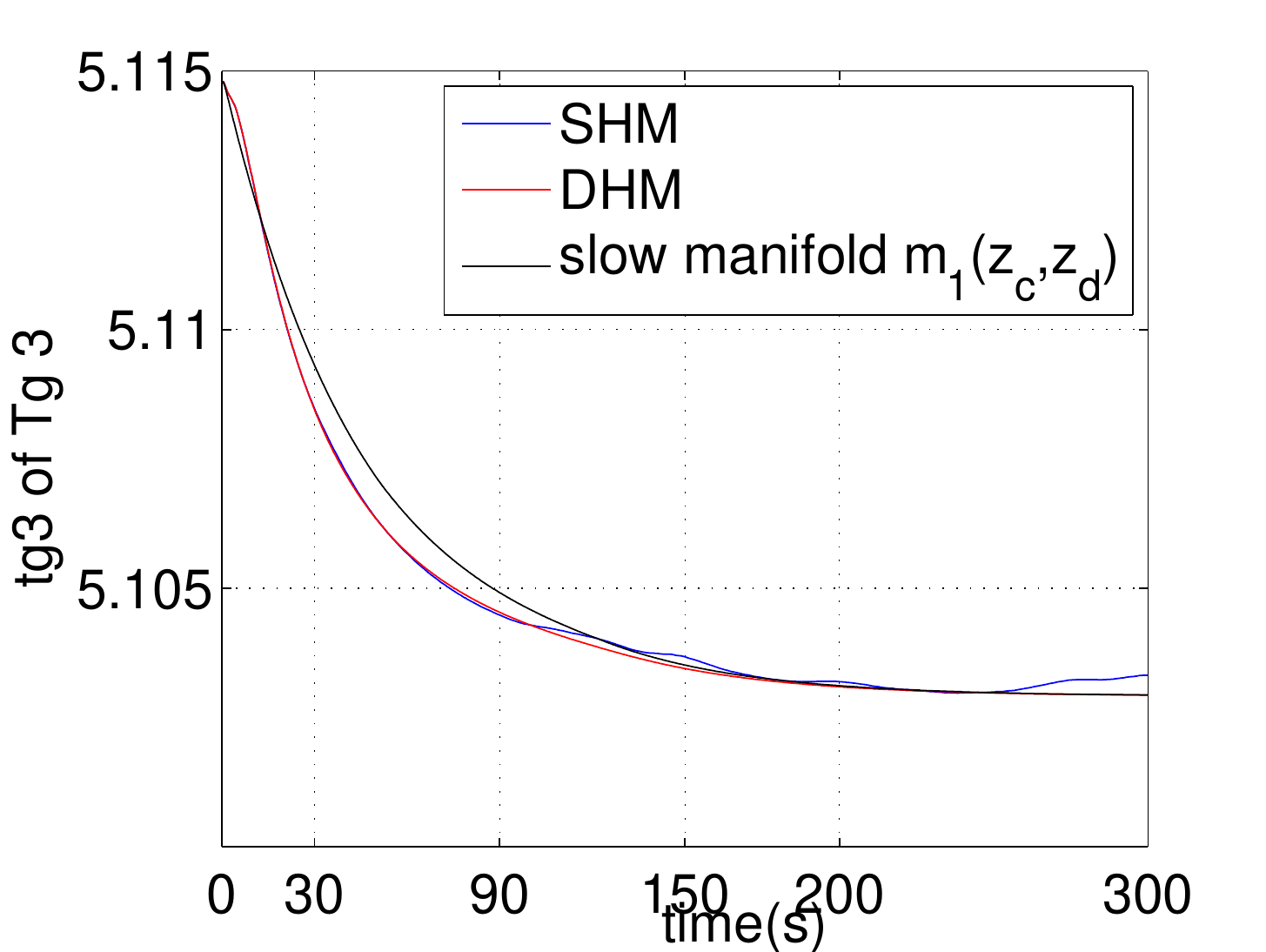}
\end{minipage}
\begin{minipage}[t]{0.5\linewidth}
\includegraphics[width=1.8in ,keepaspectratio=true,angle=0]{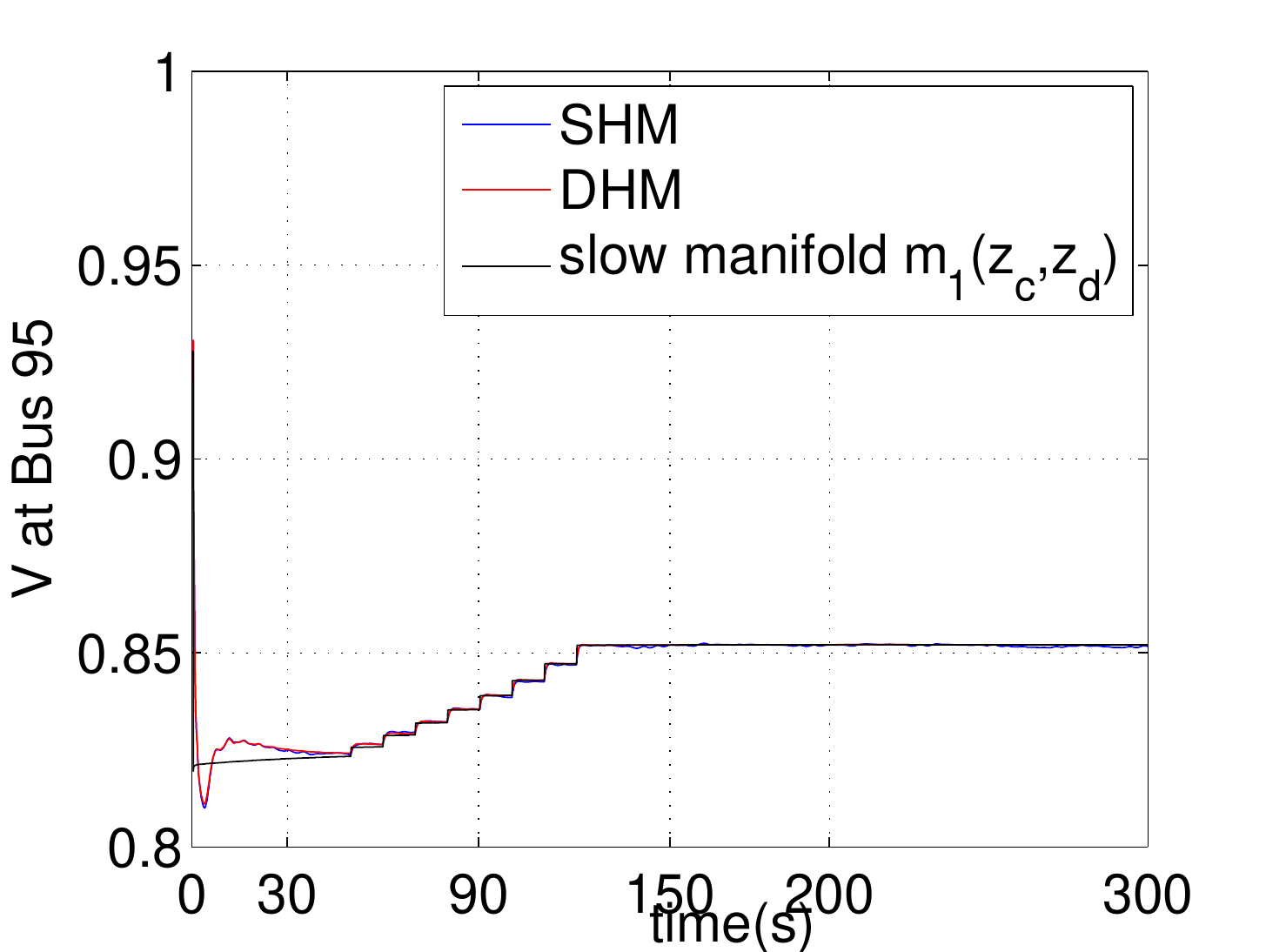}
\end{minipage}%
\begin{minipage}[t]{0.5\linewidth}
\includegraphics[width=1.8in ,keepaspectratio=true,angle=0]{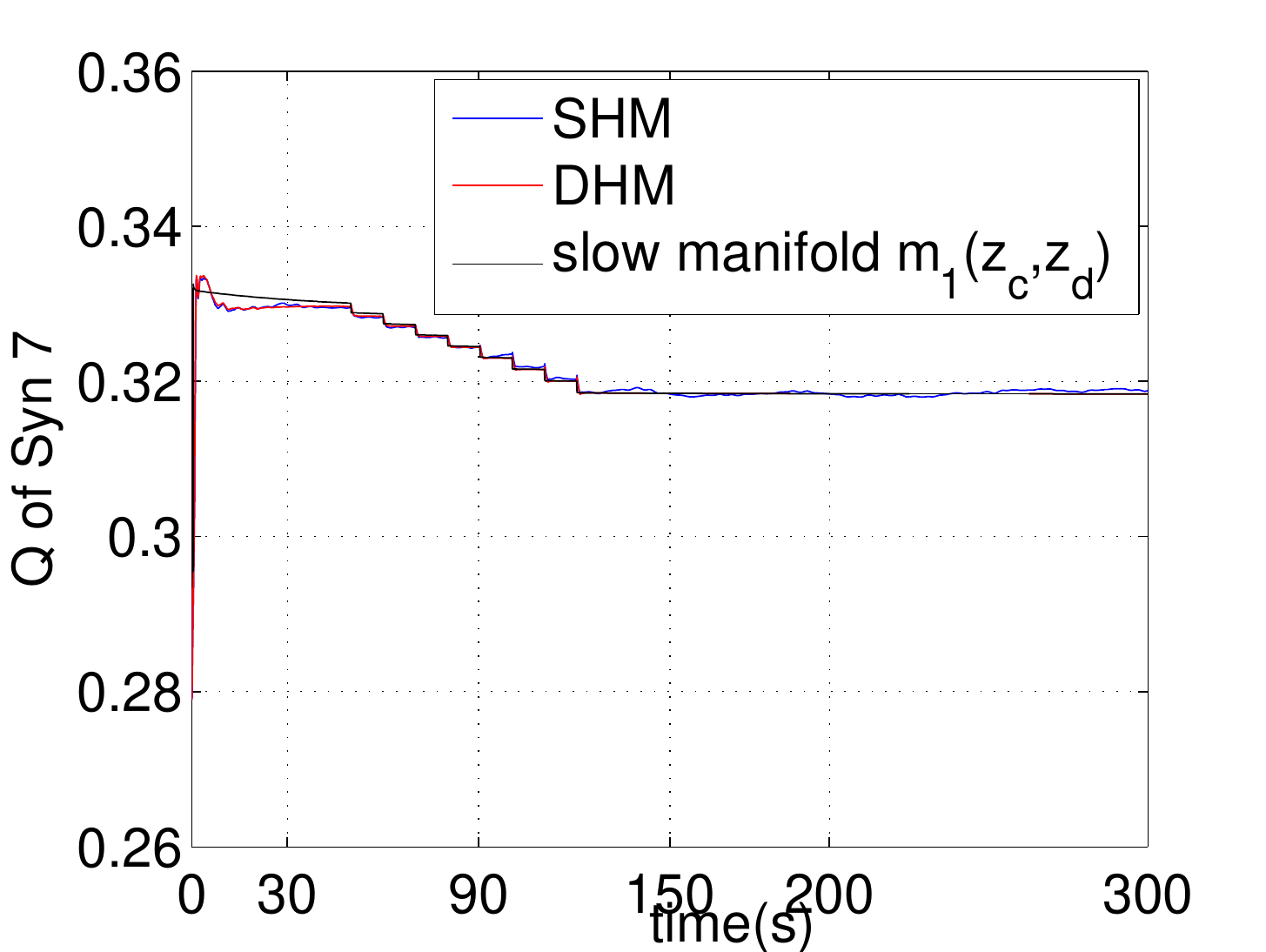}
\end{minipage}
\caption{A comparison of the trajectory of the SHM and that of the
DHM by simulating the 145-bus system.} \label{my145wind_discrete}
\end{figure}

Fig. \ref{my145wind_discrete} presents a comparison of the
trajectory of the SHM and that of the DHM for which the quasi
steady-state (QSS) model \cite{Cutsem:book} is implemented to obtain
the slow manifolds of the DHM. Observe that the trajectories of the
SHM always keep close to those of the DHM despite the changing of
discrete variables, and both models give the same stability
assessments that the system is stable in the long-term time scale.
%Based on the analytical results derived in Section \ref{subsectiontheory},
Clearly, all sufficient conditions of \textit {Theorem 2} are
satisfied, then the conclusions of \textit{Theorem 2} hold.
Particularly, Fig. \ref{my145wind_discrete} shows that the DHM does
not encounter the singularity and its slow manifold is stable. In
addition, the trajectory of the DHM evolves along
$\bm{m_1^\star}(\bm{z_c},\bm{z_d},\ee)$ which is an
$\ee$-neighborhood of the slow manifold. This illustrates the
results of \textit{Theorem 2}.

Concerning the computational efficiency, the SHM takes 137.118s to
complete the simulation, whereas the DHM only consumes 57.913s. Note
that several trajectories of the SHM may be required to evaluate the
stability in critical cases. But, the time needed to simulate one
trajectory (of the SHM) can be more than twice as that required by
the DHM.

From this example, we observe that the DHM can provide an accurate
trajectory approximation and stability assessments for the SHM with
far less simulation time, provided that the proposed mild conditions
are satisfied.

\section{necessity of stochastic model}\label{sectionnecessity}

A comprehensive theoretical framework has been developed to
approximate the SHM by the DHM. Specifically, if all sufficient
conditions of \textit{Theorem 2} are satisfied, then the DHM can
provide an accurate trajectory approximation and stability
assessments for the SHM with much less simulation time. In the
section, we further present several examples in critical cases that
the DHM fails to provide a satisfactory approximation. The causes
for such failure are investigated in the nonlinear system framework
and are shown to correspond to a violation of the proposed
sufficient conditions. This discussion complements the proposed
framework and methodology and also highlights the necessity of the
stochastic model when performing the stability analysis for the
power system with significant wind power generations, especially for
the system that operates close to the stability boundary. Given the
importance of such sufficient conditions, it is imperative to
develop efficient numerical algorithms to check these conditions in
the near future.

\subsection{Numerical Example I}

This example is a modified IEEE 14-bus system. {The order V and order
VI models of GENs are employed.} A Weibull-distributed
wind source drives a DFIG at Bus 2, and 3 GENs are equipped with
AVRs and TGs. In addition, 3 exponential recovery loads (ERLs) are
at Bus 9, 10, and 14, respectively. An OXL is installed for GEN 1,
and 3 discrete LTCs are at Bus 4-9, Bus 12-13 and Bus 2-4,
respectively, the initial time delays of which are 30s and fixed
tapping delays are 10s. At 1s, three lines at Bus 6-13, Bus 7-9, and
Bus 6-11 trip. We refer the reader to Appendix \ref{appendix2-1} for
the parameter values.

\begin{figure}[!ht]
\centering
\begin{minipage}[t]{0.5\linewidth}
\includegraphics[width=1.8in ,keepaspectratio=true,angle=0]{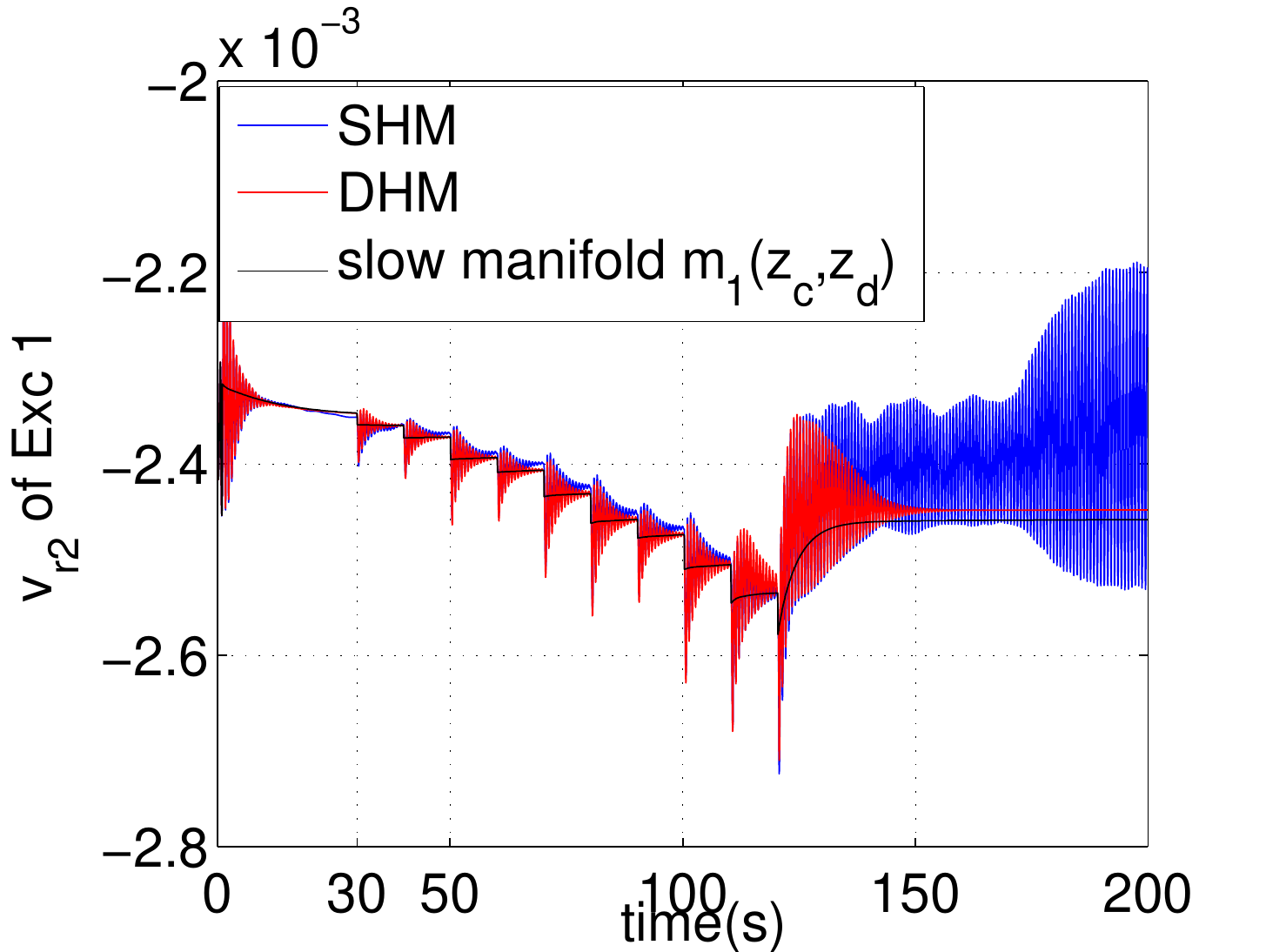}
\end{minipage}%
\begin{minipage}[t]{0.5\linewidth}
\includegraphics[width=1.8in ,keepaspectratio=true,angle=0]{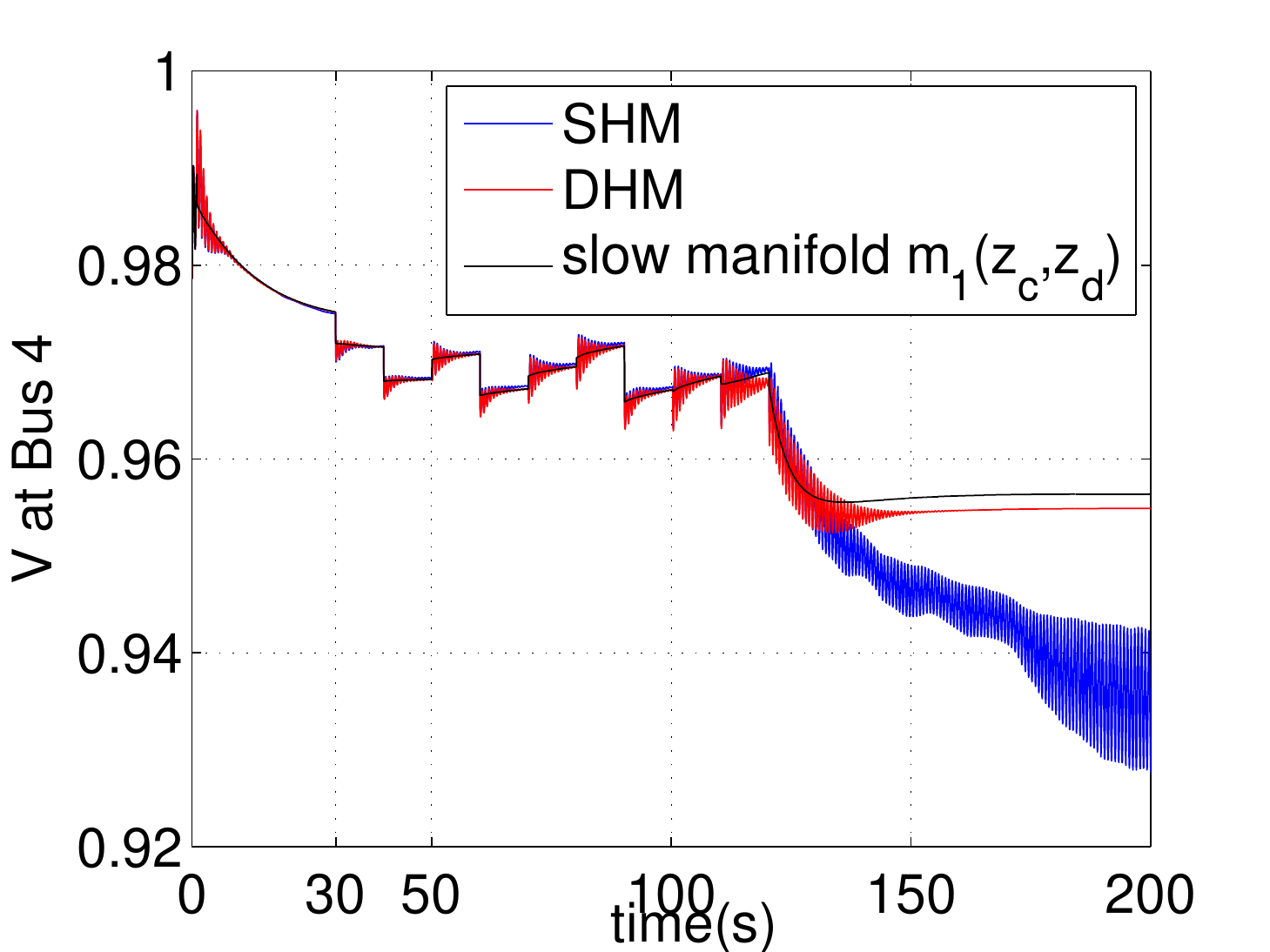}
\end{minipage}
\begin{minipage}[t]{0.5\linewidth}
\includegraphics[width=1.8in ,keepaspectratio=true,angle=0]{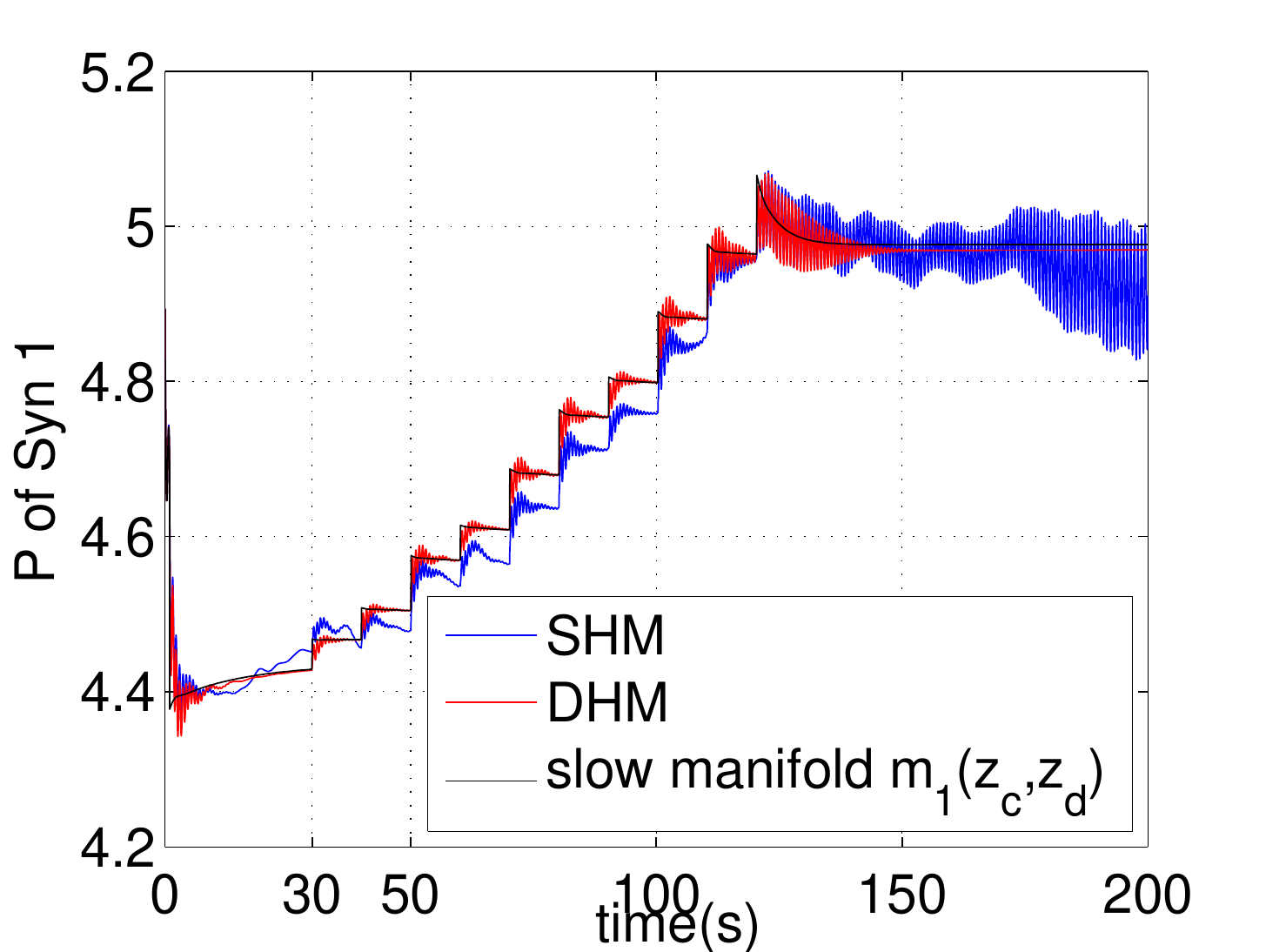}
\end{minipage}%
\begin{minipage}[t]{0.5\linewidth}
\includegraphics[width=1.8in ,keepaspectratio=true,angle=0]{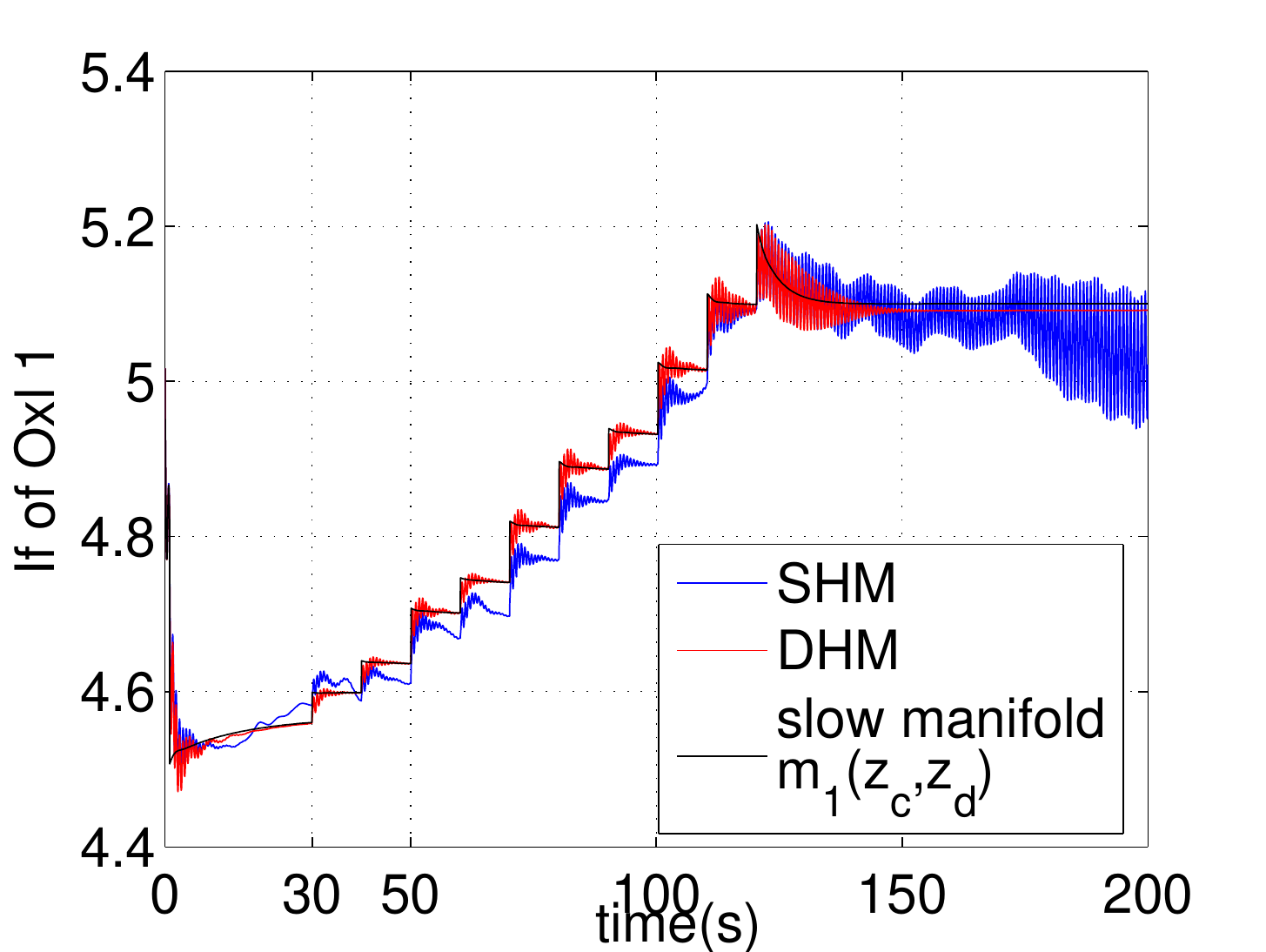}
\end{minipage}
\caption{A comparison between the trajectory of the SHM and that of
the DHM using the 14-bus system. The DHM fails to capture the
instability of the SHM.}\label{my14trywind1}
\end{figure}

A comparison between the trajectory of the DHM and that of the SHM
is shown in Fig. \ref{my14trywind1}. The slow manifold of the DHM
acquired from the QSS model is also illustrated. The DHM converges
to a long-term stable equilibrium point (SEP) with all voltages in
the nominal range, which shows that the DHM is long-term stable.
But, the sample path of the SHM suffers from a voltage collapse. So,
the DHM fails to provide a stability assessment agreeing with the
SHM.

\begin{figure}[!ht]
\centering
\begin{minipage}[t]{0.5\linewidth}
\includegraphics[width=1.8in ,keepaspectratio=true,angle=0]{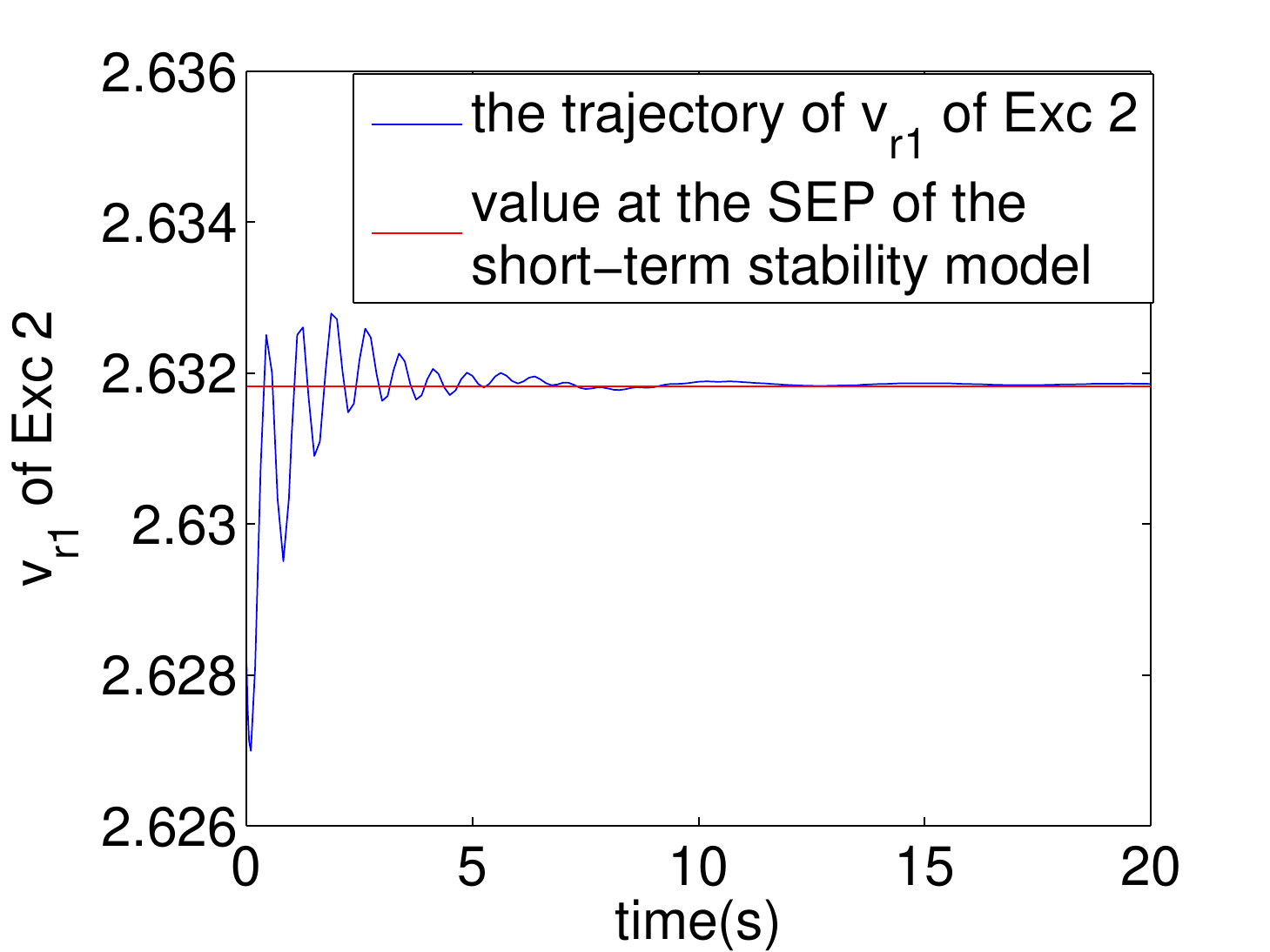}
\end{minipage}%
\begin{minipage}[t]{0.5\linewidth}
\includegraphics[width=1.8in ,keepaspectratio=true,angle=0]{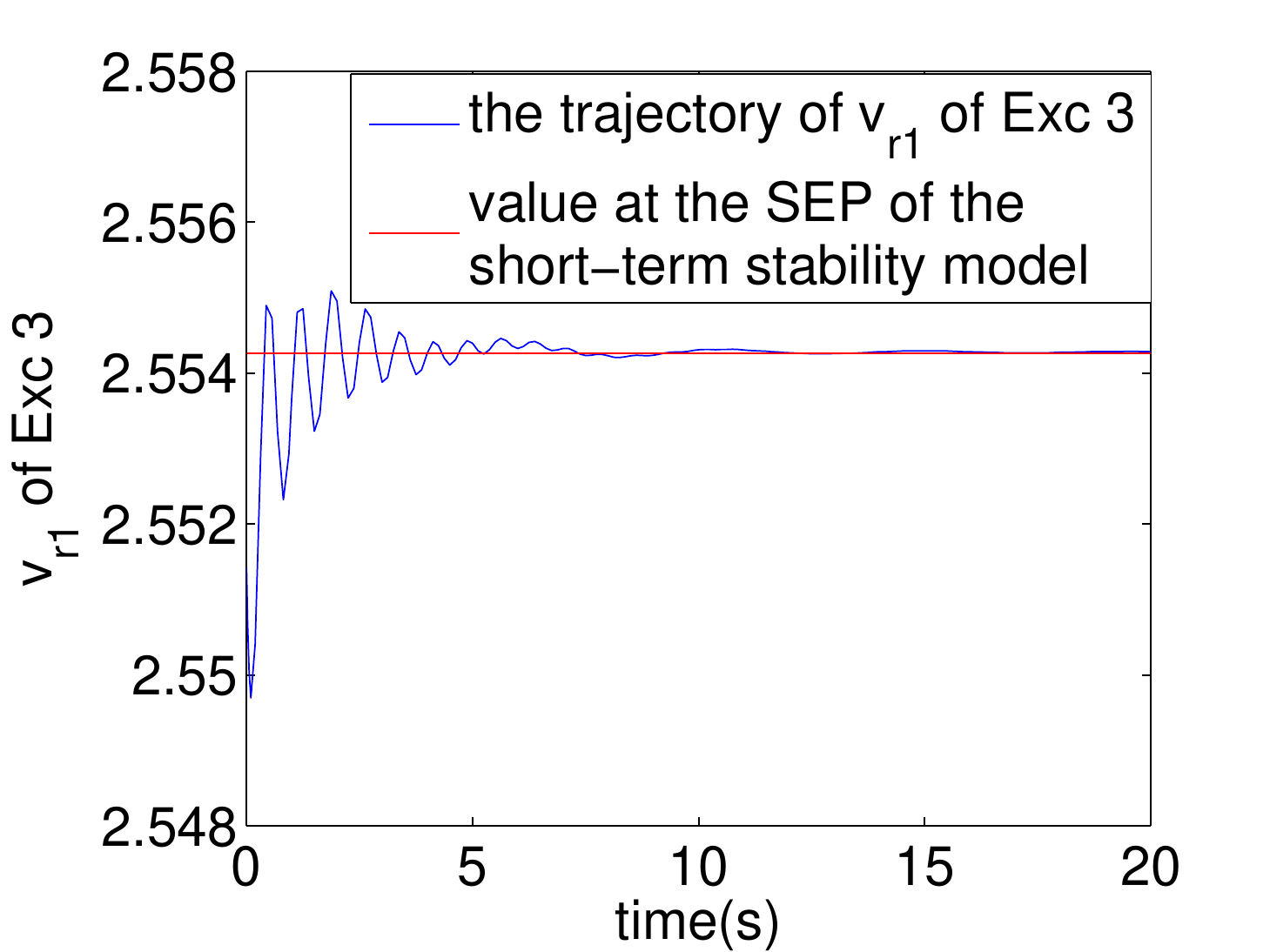}
\end{minipage}
\caption{The trajectories of two fast variables in the short-term
stability model when $\bm{z_d}$ change at 110s. The trajectory
starting from the state of DHM converges to the SEP of the
corresponding short-term stability model which shows that the
condition of consistent attraction is
satisfied.}\label{my14trywind1_fastmodel-110}
\end{figure}

The failure of the DHM is caused by a violation of condition (iii),
i.e., \textit{the condition of consistent attraction}, in
\textit{Theorem 2}. When the discrete variables (i.e., the ratios of
LTCs) change at 120s, the state of the DHM lies outside the
stability region of the corresponding short-term stability model. To
show this, the following simulations are conducted similar to the
approach in \cite{Wangxz:CAS}. When discrete variables jump at 110s,
the trajectories of two fast variables of the corresponding
short-term stability model starting from the state of the DHM are
shown in Fig. \ref{my14trywind1_fastmodel-110}. Observe that the
trajectories converge to the SEP of the corresponding short-term
stability model which shows that \textit{the condition of consistent
attraction} is satisfied at this time.

\begin{figure}[!ht]
\centering
\begin{minipage}[t]{0.5\linewidth}
\includegraphics[width=1.8in ,keepaspectratio=true,angle=0]{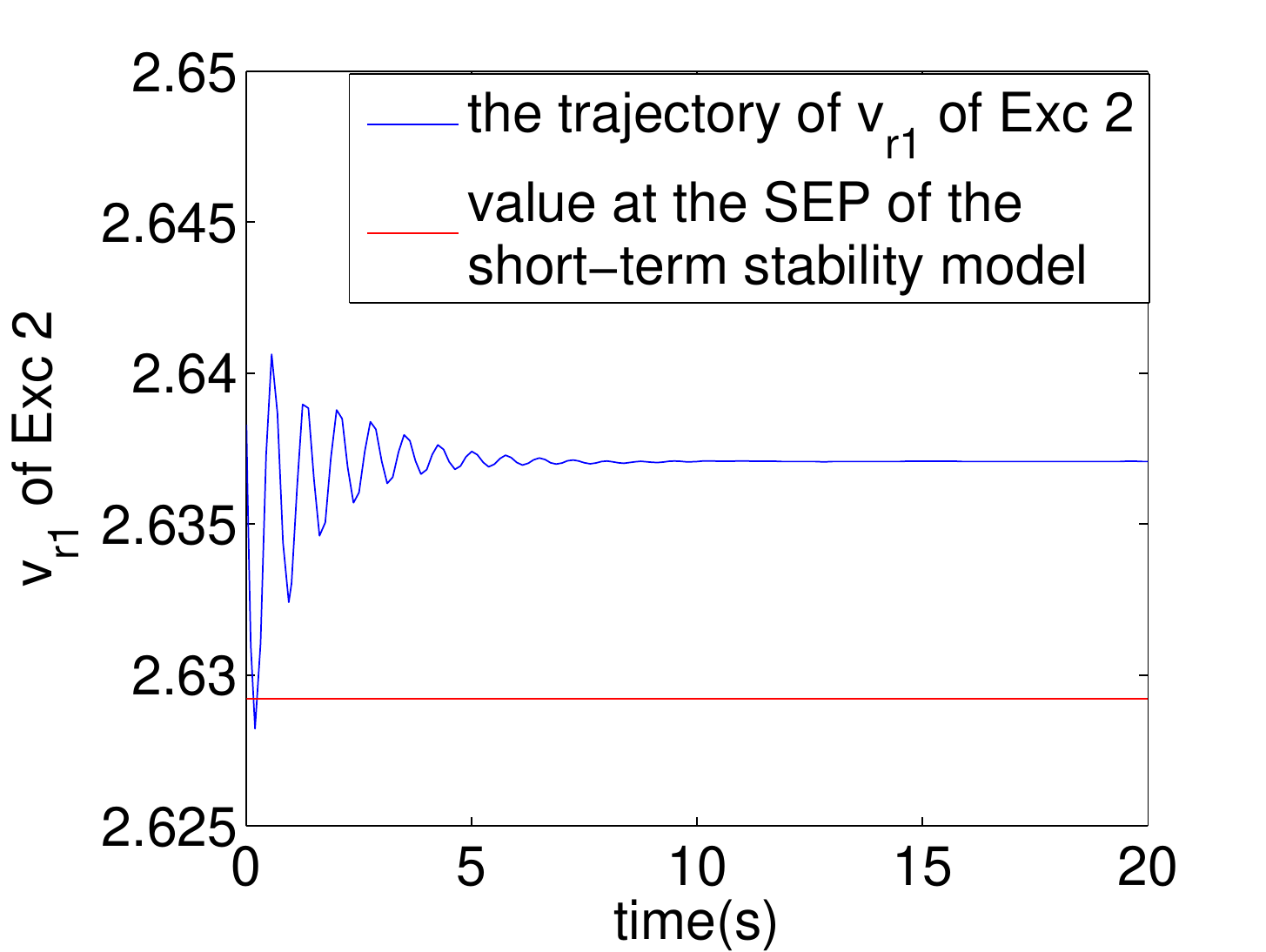}
\end{minipage}%
\begin{minipage}[t]{0.5\linewidth}
\includegraphics[width=1.8in ,keepaspectratio=true,angle=0]{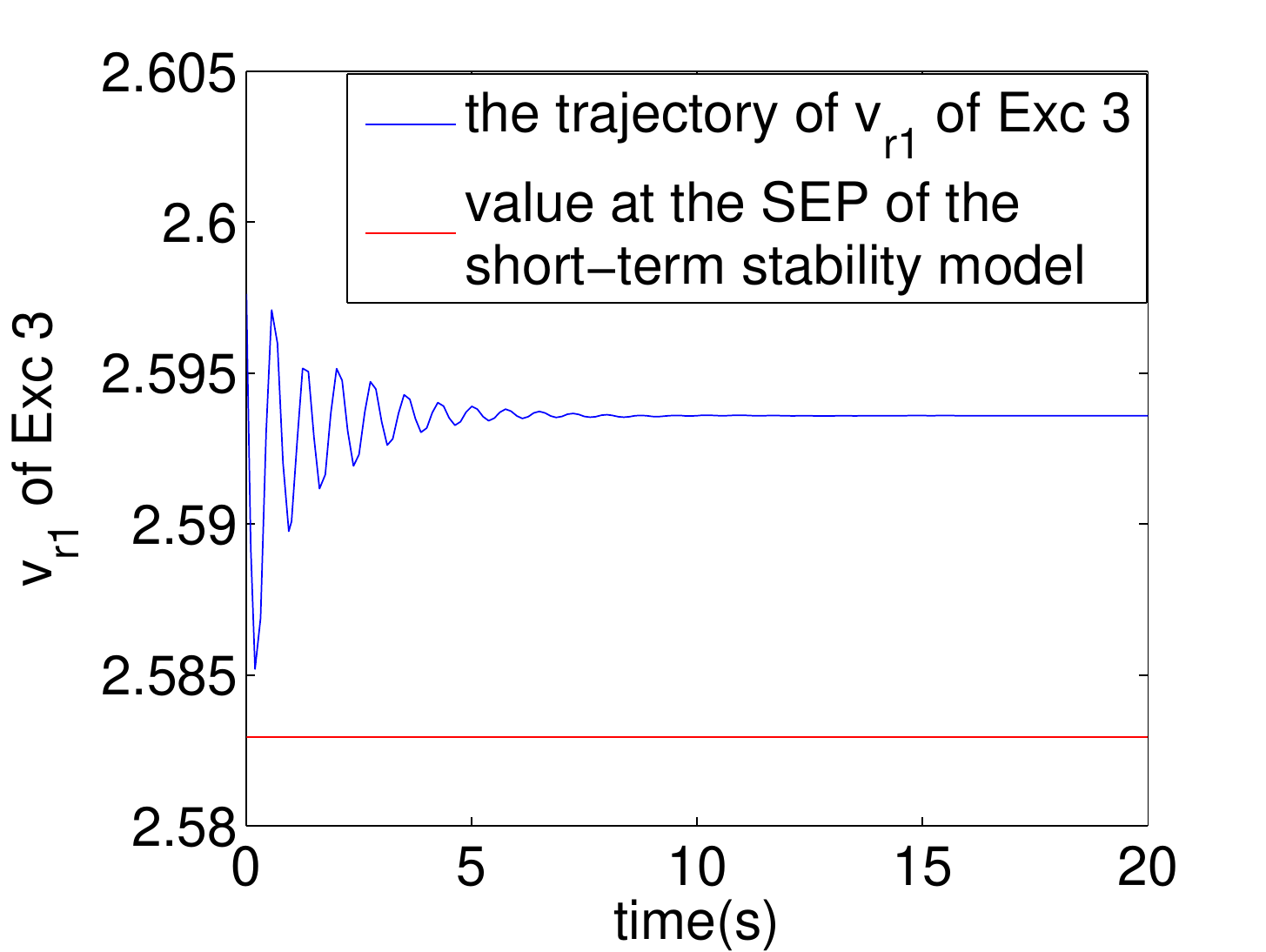}
\end{minipage}
\caption{The trajectories of two fast variables in the short-term
stability
 model when $\bm{z_d}$ change at 120s. The trajectory starting from the state
 of the DHM does not converge to the SEP of the corresponding short-term stability
 model. So, the condition of consistent attraction is
violated.}\label{my14trywind1_fastmodel-120}
\end{figure}

But, if the discrete variables jump at 120s, the trajectories of the
same two fast variables of the corresponding short-term stability
model are shown in Fig. \ref{my14trywind1_fastmodel-120}. Note that
trajectories  starting from the state of the DHM no longer converge
to the SEP of the corresponding short-term stability model, and thus
the state of the DHM lies outside the stability region of the
short-term stability model. So, condition (iii) in \textit{Theorem
2} is violated, and the DHM can fail to provide a satisfactory
approximation for the SHM, which is true in this case. Note that the
condition of consistent attraction is a sufficient but unnecessary
condition to ensure the stability of the DHM (Theorem 5-6
\cite{Wangxz:CAS}). So, the DHM is stable in this case even though
the condition of consistent contraction is violated.

\begin{figure}[!ht]
\centering
\includegraphics[width=2.5in,keepaspectratio=true,angle=0]{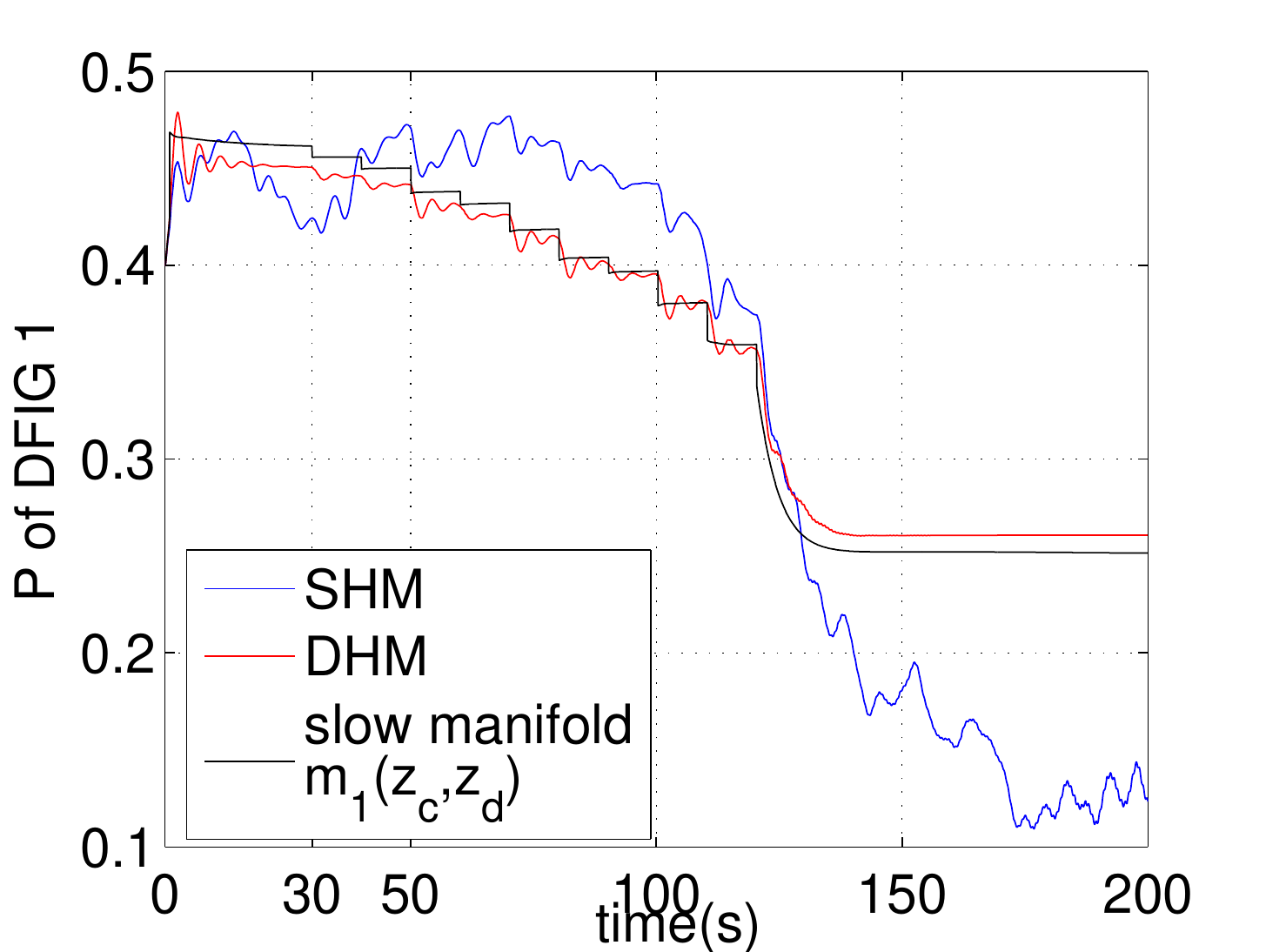}
\caption{A comparison of the real power output of DFIG in the SHM
and that in the DHM. }\label{windpower}
\end{figure}

From the viewpoint of physical mechanisms, the voltage collapse is
caused by an insufficient power support when LTCs try to restore the
load-side voltages in the long-term time scale. Immediately after
the contingency, the system can maintain the short-term stability by
the control of exciters. After that, LTCs start to work at 30s and
try to restore the load-side voltages and then the corresponding
load powers. At 112s, the OXL at GEN 1 is activated to protect the
generator from overheating and thus restrict the power support from
GEN 1. To make it even worse, the power output of DFIG at Bus 2
suddenly decreases as shown in Fig. \ref{windpower} because of a
sharp drop in the wind speed. The power imbalance between the loads
and the generators finally leads to the voltage collapse in the SHM.
In the DHM, however, the wind power does not change drastically as
shown in Fig. \ref{windpower}, since the wind speed is supposed to
be invariable. So, the DFIG at Bus 2 can provide enough power
required by the action of LTCs to maintain the voltage stability of
the DHM.

From this example, some important physical insights can be obtained.
If the wind power plays a significant role in supporting power to
maintain the stability, for example when the penetration level is
high (8.42\% in this example), then the stochastic properties of the
wind may need to be considered in the stability analysis, especially
when the system operates close to the stability boundary.

\subsection{Numerical Example II}

The second example using an IEEE 9-bus system is presented to reveal
another cause for the failure of the DHM. In the system, {the classical model of GEN is employed.} A
Weibull-distributed wind source drives a DFIG at Bus 3, and three
GENs are equipped with TGs, AVRs, and OXLs, respectively, where the
initial time delays of OXLs are 70s. In addition, three ERLs are
located at Bus 5, 6, and 8, respectively; while three discrete LTCs
are located at Bus 5-4, Bus 9-6, and Bus 2-7, respectively, the
initial time delays of which are 60s and fixed tapping delays are
10s. At 1s, a fault occurs at Bus 6 and is cleared 5 cycles later.
The parameter values are detailed in Appendix \ref{appendix2-2}.

\begin{figure}[!ht]
\centering
\begin{minipage}[t]{0.5\linewidth}
\includegraphics[width=1.8in ,keepaspectratio=true,angle=0]{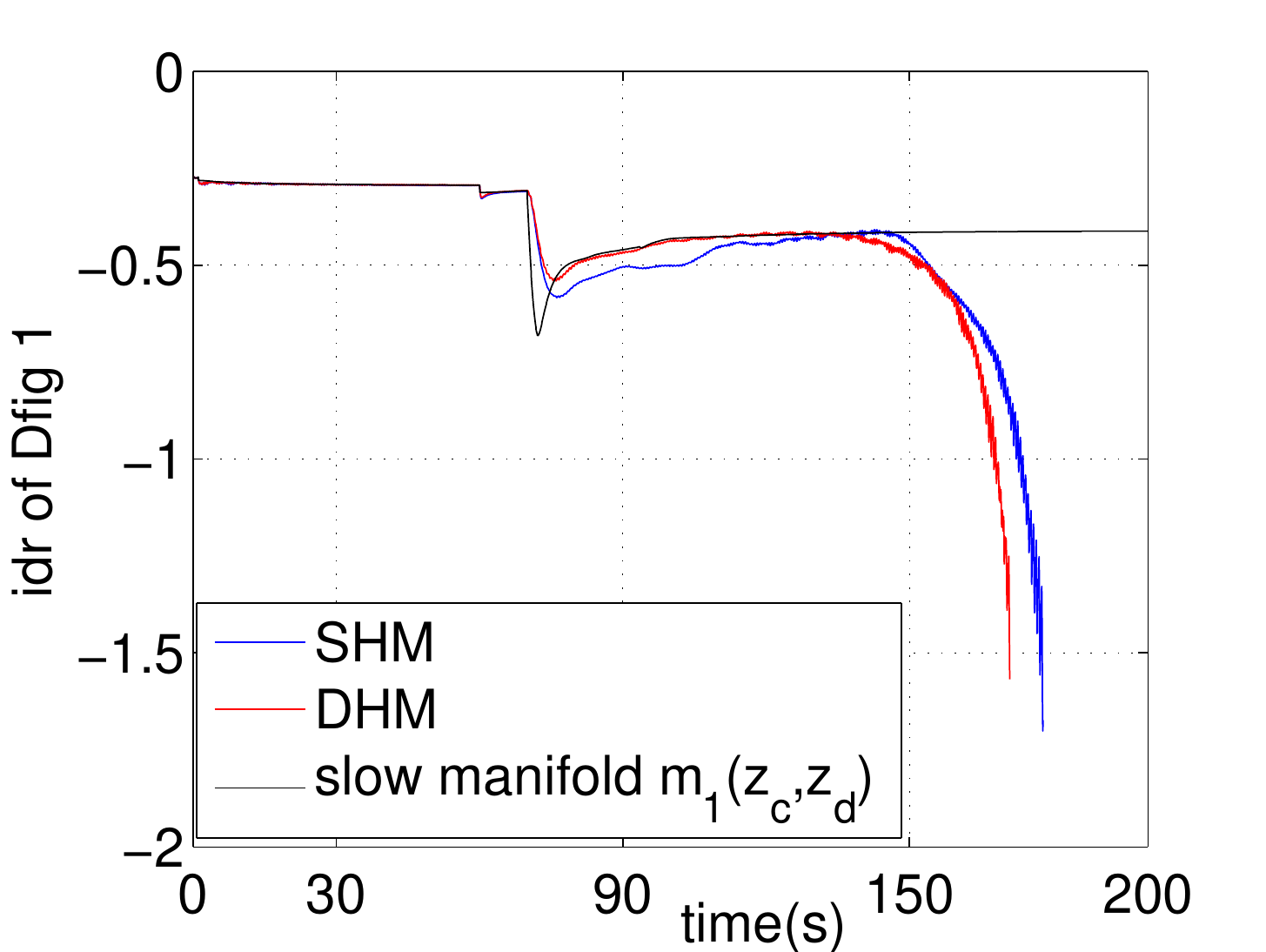}
\end{minipage}%
\begin{minipage}[t]{0.5\linewidth}
\includegraphics[width=1.8in ,keepaspectratio=true,angle=0]{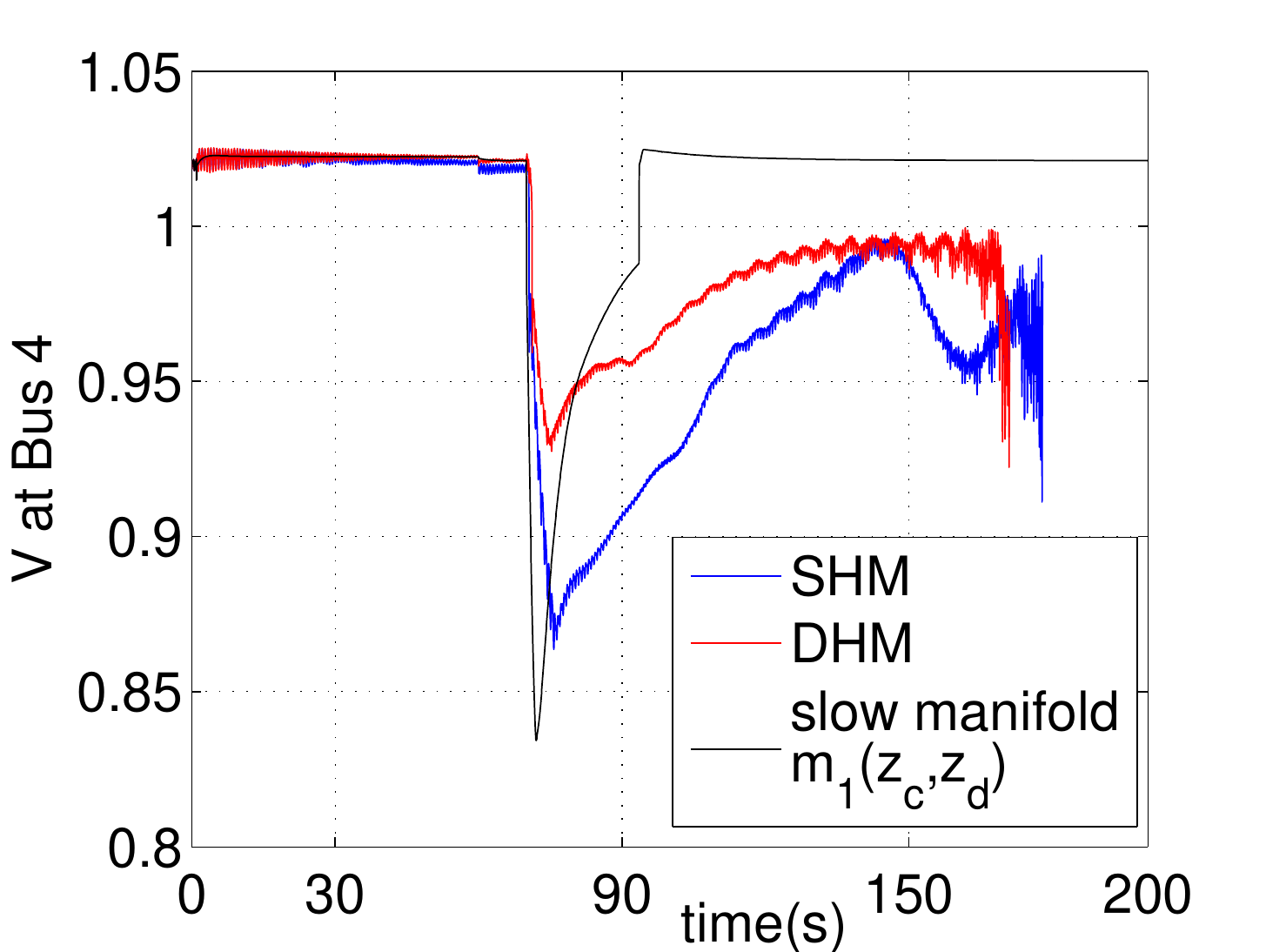}
\end{minipage}
\begin{minipage}[t]{0.5\linewidth}
\includegraphics[width=1.8in ,keepaspectratio=true,angle=0]{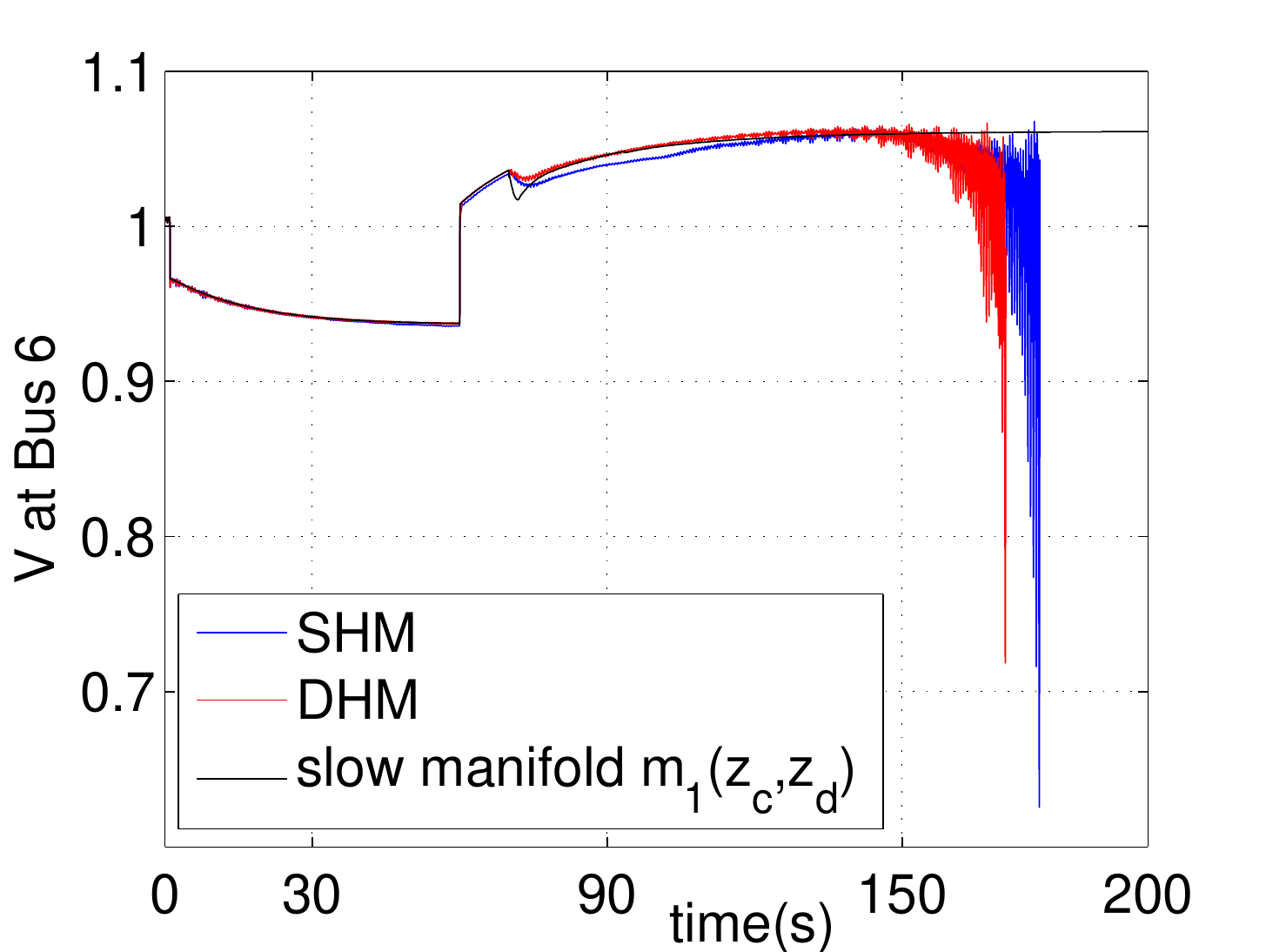}
\end{minipage}%
\begin{minipage}[t]{0.5\linewidth}
\includegraphics[width=1.8in ,keepaspectratio=true,angle=0]{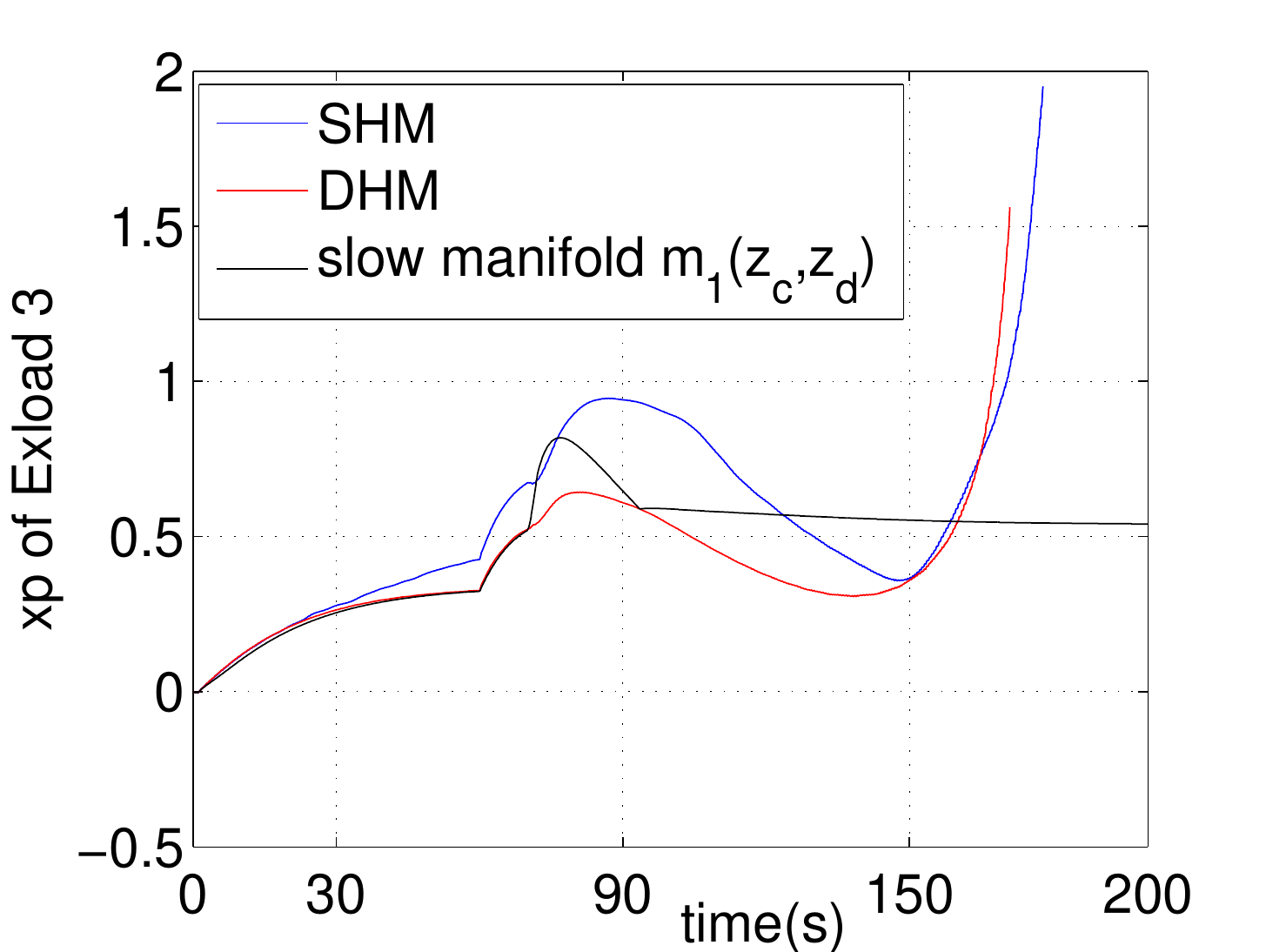}
\end{minipage}
\caption{A comparisons between the trajectories of SHM and those of
DHM. The results of \textit{Theorem 2} do not apply in this case as
condition (i) is violated.}\label{my145wind}
\end{figure}

In this case, the slow manifold $\bm{m_1(\bm{z_c},\bm{z_d})}$ is
unstable, which implies that nearby dynamics will move away from the
slow manifold. As condition (i) in \textit{Theorem 1} and
\textit{Theorem 2} is violated, neither the concentration of sample
path stated in \textit{Theorem 1} nor the trajectory relationship
described in \textit{Theorem 2} holds. The trajectory of the SHM is
not concentrated around that of the DHM, i.e., the DHM cannot
provide an accurate trajectory approximation for the SHM, but both
of them are unstable in the long-term sense.

From the perspective of physical mechanisms, the instability is
caused by the poor control of LTCs which are originally designed to
help maintain the stability. The discrete switching of LTCs makes
the slow manifold jump from the stable component of the constraint
manifold to an unstable component such that the nearby trajectories
move away. The switching events, such as LTCs and shunt
compensation, are adopted commonly as countermeasures against the
voltage instability. But, this example shows that great caution is
necessary when executing those control strategies, because
unexpected stability issues may arise, especially when more wind
power is integrated into the power grid.

\section{Concluding Remarks}\label{sectionconclusion}

{This paper proposes a comprehensive SDE-based
framework for conducting the long-term stability analysis for the
power grid with wind power generations. This framework incorporates
the discrete dynamics induced by various control devices and the
stochastic model of the wind speed with different probability
distributions. To relieve the computational burden, a DHM is
composed and can provide an accurate trajectory approximation and
correct stability assessments for the SHM under some mild sufficient
conditions. Numerical examples are further discussed to show that
the DHM can fail in some critical cases because of a violation of
the proposed sufficient conditions, which complements the proposed
SDE-based framework and also highlights the necessity of the SHM in
the stability analysis, especially if the system operates close to
the stability boundary or experiences a high variability. For the
future work, we plan to extend the present framework to the
stability analysis of power grids with various other uncertainties
and further improve the computational efficiency of the
approximation methodology using the QSS model that integrates
uncertainties.}

%The operation and control of modern power grids have to account for
%more and more uncertainty and randomness originating not only from
%the supply side (e.g., renewable energy sources) but also from the
%demand side (e.g., smart appliances). In the future, we intend to
%generalize the proposed framework and methodology to study stability
%of power grids with uncertainty/randomness resulting from different
%sources. We also plan to improve the approximation methodology by
%using the QSS model \cite{Cutsem:book}\cite{Wangxz:CAS} that can
%further reduce the simulation time. Incorporating uncertainty into
%the QSS model may be an important progress in dynamic stability
%analysis, given the wide applications of the QSS model and the
%increasing uncertainty in power grids.

\appendices
\section{Proof of Theorem 1}\label{prooftheorem1}

\textit{Proof:} Conditions (i) and (ii) {ensure}
that all conditions of \textit{Theorem 1} \cite{Wangxz:sde1} are
satisfied for each fixed $\bm{z_d}(k)$, $k=0,1,...,N$. So, the
conclusions of \textit{Theorem 1} \cite{Wangxz:sde1} are valid for
each continuous system of the SHM with fixed $\bm{z_d}(k)$. Then,
there exist $\ee_0^k>0$, $h_0^k>0$, $\delta_0^k>0$, and a time
$\tilde{\tau}_k$ of order $\ee|\mbox{log}h|$ such that whenever
$\delta\leq\delta_0^k$, the following inequality
\begin{eqnarray}\label{theorem2}
&&\mathbb{P}\{\exists \tau\in[\tilde{\tau}_k,\tau_k):
(\bm{z_c}^k(\tau),\bm{\bar{x}}^k(\tau),\bm{z_d}(k))\notin M(h)\}\nonumber\\
&&\leq
C_{n_{z_c},n_x}(\tau,\ee)e^{\frac{-h^2}{2\sigma^2}(1-O(h)-O(\ee))}
\end{eqnarray}
holds for all $\ee\leq\ee_0^k$, $h\leq h_0^k$, $k\in[0,1...N]$, on
$[\tilde{\tau}_k,\tau_{k+1})$ for $k\in[0,1,...N-1]$ or on
$[\tilde{\tau}_k,\infty]$ for $k=N$. Here,
$(\bm{z_c}^k(\tau),\bm{\bar{x}}^k(\tau),\bm{z_d}(k))$ is the
solution of each continuous system (\ref{sde power slow})-(\ref{sde
power fast}) of the SHM for fixed $\bm{z_d}(k)$ with initial
condition $(\bm{z_c}^k(0),\bm{\bar{x}}^k(0),\bm{z_d}(k))$.

Let $\ee_0=\mbox{min}(\ee_0^0,\ee_0^1,...\ee_0^N)$,
$h_0=\mbox{min}(h_0^0,h_0^1,...h_0^N)$, and
$\delta_0=\mbox{min}(\delta_0^0,\delta_0^1,...\delta_0^N)$. Then,
for
$\tau\in\Pi=\cup_{i=1}^{N-1}[\tilde{\tau}_i,\tau_i)\cup[\tilde{\tau}_N,\infty]$,
the following inequality
\begin{eqnarray}\label{theorem2}
&&\mathbb{P}\{\exists \tau\in \Pi:~ (\bm{z_c}(\tau),\bm{\bar{x}}(\tau),\bm{z_d})\notin M(h)\}\nonumber\\
&&\leq
C_{n_{z_c},n_x}(\tau,\ee)e^{\frac{-h^2}{2\sigma^2}(1-O(h)-O(\ee))}
\end{eqnarray}
holds for all $\ee\leq\ee_0$, $h\leq h_0$. This completes the proof.

\section{Proof of Theorem 2}\label{prooftheorem2}

Conditions (i)-(iii) ensure that all conditions of \textit{Theorem
2} \cite{Wangxz:sde1} are satisfied for each fixed $\bm{z_d}(k)$,
$k=0,1,...,N$. So, the conclusions of \textit{Theorem 2}
\cite{Wangxz:sde1} are valid for each continuous system of the SHM
with fixed $\bm{z_d}(k)$. So, there {exist}
$\ee_0^k>0$, $\delta_0^k>0$, a time $\tilde{\tau}_k$ of order
$\ee|\mbox{log}h|$, and $\bar{\tau}_k$ such that whenever
$\delta\leq\delta_0^k$ for all
$\tau\in[\tilde{\tau}_k,\bar{\tau}_k]$, the following estimates
\begin{eqnarray}
|\bm{\bar{x}}^k(\tau)-\bm{\bar{x}_D}^k(\tau)| &=& O(\sigma),\\
|\bm{z_{c}}^k(\tau)-\bm{z_{cD}}^k(\tau)| &=& O(\sigma\sqrt{\ee}),
\end{eqnarray}
hold for all $\ee\in(0,\ee_0^k)$, $0\le k\le N$, Here, $(
\bm{z_c}^k(\tau), \bm{\bar{x}}^k(\tau),$ $\bm{z_d}(k) )$ is the
solution of each continuous system (\ref{sde power slow})-(\ref{sde
power fast}) of the SHM, and
$(\bm{z_{cD}}^k(\tau),\bm{\bar{x}_D}^k(\tau),\bm{z_d}(k))$ is the
solution of each continuous system (\ref{det power slow})-(\ref{det
power fast}) of the DHM for fixed $\bm{z_d}(k)$.

Let $\ee_0=\mbox{min}(\ee_0^0,\ee_0^1,\cdots,\ee_0^N)$,
$h_0=\mbox{min}(h_0^0,h_0^1,\cdots,h_0^N)$, and
$\delta_0=\mbox{min}(\delta_0^0,\delta_0^1,\cdots,\delta_0^N)$.
Similar to \textit{Theorem 1}, one can show that for all
$\tau\in\cup_{i=1}^{N}[\tilde{\tau}_i,\bar{\tau}_i]$, the following
estimates
\begin{eqnarray}
|\bm{\bar{x}}(\tau)-\bm{\bar{x}_D}(\tau)|&=&O(\sigma)\label{corollary3_1}\\
|\bm{z_{c}}(\tau)-\bm{z_{cD}}(\tau)|&=&O(\sigma\sqrt{\ee})\label{corollary3_2}
\end{eqnarray}
hold for all $\ee\in(0,\ee_0)$. The proof of the theorem is
completed.

\section{Parameter Values of Numerical Example I}\label{appendix2-1}

The system is modified from the 14-bus test case in PSAT-2.1.6. The
GEN at Bus 2 is replaced by DFIG. The parameter values are given in
Table \ref{appedixtable1}-\ref{appedixtable5}.

\begin{table}[!ht]
\centering \caption{Doubly-fed induction generator parameter
values}\label{appedixtable1}
\begin{tabular}{|c|c|}
\hline
Parameter&Value\\
\hline
stator resistance $r_s$&$0.01$p.u.\\
\hline
stator reactance $x_s$ &$0.1$p.u.\\
\hline
rotor resistance $r_r$&$0.01$p.u.\\
\hline
rotor reactance $x_r$&$0.08$p.u.\\
\hline
magnetizing reactance $x_\mu$&$3$p.u.\\
\hline
rotor inertia $H_m$&$3$KWs/KVA\\
\hline
pitch control gain $K_p$&$10$\\
\hline
pitch control time constant $T_p$&$3$s\\
\hline
voltage control gain $K_v$&10\\
\hline
power control time constant $T_\ee$&$0.01$s\\
\hline
rotor radius $R$&$75$m\\
\hline
number of poles $n_p$&$4$\\
\hline
number of blades $n_b$&$3$\\
\hline
gear box ratio $\eta_{GB}$&$0.0112$\\%$0.01123596$
\hline
maximum active power $p^{max}$&$2$p.u.\\
\hline
minimum active power $p^{min}$&$-1$p.u.\\
\hline
maximum reactive power $q^{max}$&$2$p.u.\\
\hline
minimum reactive power $q^{min}$&$-1$p.u.\\
\hline
number of machines $n_g$&$1$\\
\hline
\end{tabular}
\end{table}

\begin{table}[!ht]
\centering \caption{Turbine governor parameter
values}\label{appedixtable2}
\begin{tabular}{|c|c|}
\hline
Parameter&Value\\
\hline
reference speed $\omega^{0}_{ref}$&$1$p.u.\\
\hline
droop $R$&$0.02$p.u.\\
\hline
maximum turbine output $p^{max}$&$1.2$p.u.\\
\hline
minimum turbine output $p^{min}$&$0$p.u.\\
\hline
governor time constant $T_s$&$0.1$s\\
\hline
servo time constant $T_c$&$0.45$s\\
\hline
transient gain time constant $T_3$&$0$s\\
\hline
power fraction time constant $T_4$&$12$s\\
\hline
reheat time constant $T_5$&$50$s\\
\hline
\end{tabular}
\end{table}

\begin{table}[!ht]
\centering \caption{Load tap changer parameter values for the ones
at bus 4-9, bus 12-13 and bus 2-4}\label{appedixtable3}
\begin{tabular}{|c|c|}
\hline
Parameter&Value\\
\hline
the reference voltage $v_0$&$1.005$, $1.01$, $0.995$ \\
%&for LTC at Bus 4-9, Bus 12-13\\
%&and Bus 2-4 respectively\\
\hline
half of the deadband $d$&$0.005$, $0.1$, $0.025$ p.u.\\
%&for LTC at Bus 4-9, Bus 12-13\\
%&and Bus 2-4 respectively\\
\hline
tap step $r$&$0.025$\\
\hline
upper tap limit $r^{max}$&$1.2$\\
\hline
lower tap limit $r^{min}$&$0.7$\\
\hline
the initial time delay $\triangle{T_0}$&$30$s\\
\hline
the sequential time delay $\triangle{T_k}$&$10$s\\
\hline
\end{tabular}
\end{table}

\begin{table}[!ht]
\centering \caption{Exponential recovery load parameter
values}{\label{appedixtable4}}
\begin{tabular}{|c|c|}
\hline
Parameter&Value\\
\hline
active power percentage $k_p$&$100\%$\\
\hline
reactive power percentage $k_q$&$100\%$\\
\hline
active power time constant $T_p$&$10$s\\
\hline
reactive power time constant $T_q$&$10$s\\
\hline
static active power exponent $\alpha_s$&$1$\\
\hline
dynamic active power exponent $\alpha_t$&$1.5$ for the load at Bus 9\\
&$5$ for the others\\
\hline
static reactive power exponent $\beta_s$&$2$\\
\hline
dynamic reactive power exponent $\beta_t$&$2.5$ for the load at Bus 9\\
&$10$ for the others\\
\hline
\end{tabular}
\end{table}

\begin{table}[!ht]
\centering \caption{Over excitation limiter parameter
values}\label{appedixtable5}
\begin{tabular}{|c|c|}
\hline
Parameter&Value\\
\hline
maximum field current $i_f^{lim}$&$5.1$p.u.\\
\hline
integrator time constant $T_0$&$12$s\\
\hline
maximum output signal $v_{\mbox{oxl}}$&$100$p.u.\\
\hline
\end{tabular}
\end{table}

\section{Parameter Values of Numerical Example II}\label{appendix2-2}

The system is modified from the 9-bus test system in PSAT-2.1.6.
There is a DIFG at Bus 3. The parameters of the DFIG are the same as
those for Numerical Example I in Table \ref{appedixtable1}. The
parameters of other devices are shown in Table
\ref{appedixtable6}-\ref{appedixtable9}.

\begin{table}[!ht]
\centering \caption{Turbine governor parameter
values}\label{appedixtable6}
\begin{tabular}{|c|c|}
\hline
Parameter&Value\\
\hline
reference speed $\omega^{0}_{ref}$&$1$p.u.\\
\hline
droop $R$&$0.02$p.u.\\
\hline
maximum turbine output $p^{max}$&$2$p.u.\\
\hline
minimum turbine output $p^{min}$&$0.3$p.u.\\
\hline
governor time constant $T_s$&$0.1$s\\
\hline
servo time constant $T_c$&$0.45$s\\
\hline
transient gain time constant $T_3$&$0$s\\
\hline
power fraction time constant $T_4$&$12$s\\
\hline
reheat time constant $T_5$&$50$s\\
\hline
\end{tabular}
\end{table}

\begin{table}[!ht]
\centering \caption{Exponential recovery load parameter
values}{\label{appedixtable7}}
\begin{tabular}{|c|c|}
\hline
Parameter&Value\\
\hline
active power percentage $k_p$&$40\%$\\
\hline
reactive power percentage $k_q$&$40\%$\\
\hline
active power time constant $T_p$&$10$s\\
\hline
reactive power time constant $T_q$&$10$s\\
\hline
static active power exponent $\alpha_s$&$1$\\
\hline
dynamic active power exponent $\alpha_t$&$10$ for the load at Bus 4\\
&$5$ for the others\\
\hline
static reactive power exponent $\beta_s$&$2$\\
\hline
dynamic reactive power exponent $\beta_t$&$20$ for the load at Bus 4\\
&$10$ for the others\\
\hline
\end{tabular}
\end{table}

\begin{table}[!ht]
\centering \caption{Load tap changer parameter values for the ones
at bus 5-4, bus 9-6, and bus 2-7}\label{appedixtable8}
\begin{tabular}{|c|c|}
\hline
Parameter&Value\\
\hline
the reference voltage $v_0$&$1.005$, $1.005$, $1.02$ \\
\hline
half of the deadband $d$&$0.025$, $0.025$, $0.04$ p.u.\\
\hline
tap step $r$&$0.12$\\
\hline
upper tap limit $r^{max}$&$1.1$\\
\hline
lower tap limit $r^{min}$&$0.9$\\
\hline
the initial time delay $\triangle{T_0}$&$60$s\\
\hline
the sequential time delay $\triangle{T_k}$&$10$s\\
\hline
\end{tabular}
\end{table}

\begin{table}[!ht]
\centering \caption{Over excitation limiter parameter
values}\label{appedixtable9}
\begin{tabular}{|c|c|}
\hline
Parameter&Value\\
\hline
maximum field current $i_f^{lim}$&$2.02$, $1.3$, $1.32$p.u. \\
\hline
integrator time constant $T_0$&$10$s for GEN 1-2\\
&$30$s for GEN 3\\
\hline
maximum output signal $v_{\mbox{oxl}}$&$100$p.u.\\
\hline
\end{tabular}
\end{table}

\ifCLASSOPTIONcaptionsoff
  \newpage
\fi

\begin{IEEEbiography}[{\includegraphics[width=1.2in,clip,keepaspectratio,angle=270]{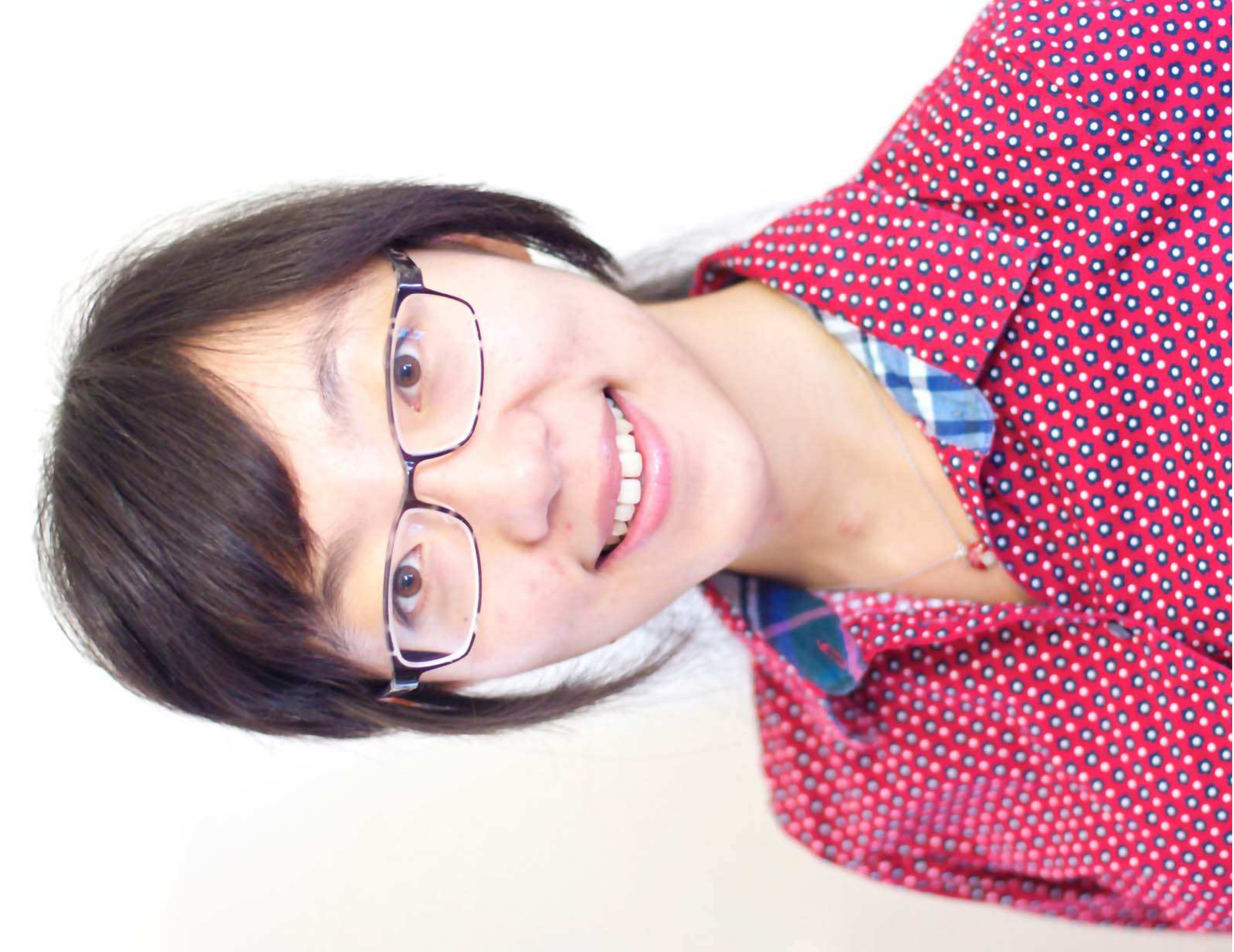}}]{Xiaozhe Wang}
(S'13-M'15) is currently an Assistant Professor in the department of Electrical and Computer Engineering at McGill University. She received the Ph.D. degree and the M. Eng in the School of Electrical and Computer Engineering from Cornell University, Ithaca, NY, USA, in 2015 and 2011, respectively, and the B.S. degree in Information Science \& Electronic Engineering from Zhejiang University, Zhejiang, China, in 2010. She was a Research Aid Intern at Argonne National Laboratory, Argonne, IL, USA, in 2014.
\end{IEEEbiography}

\begin{IEEEbiography}[{\includegraphics[width=1in,clip,keepaspectratio,angle=0]{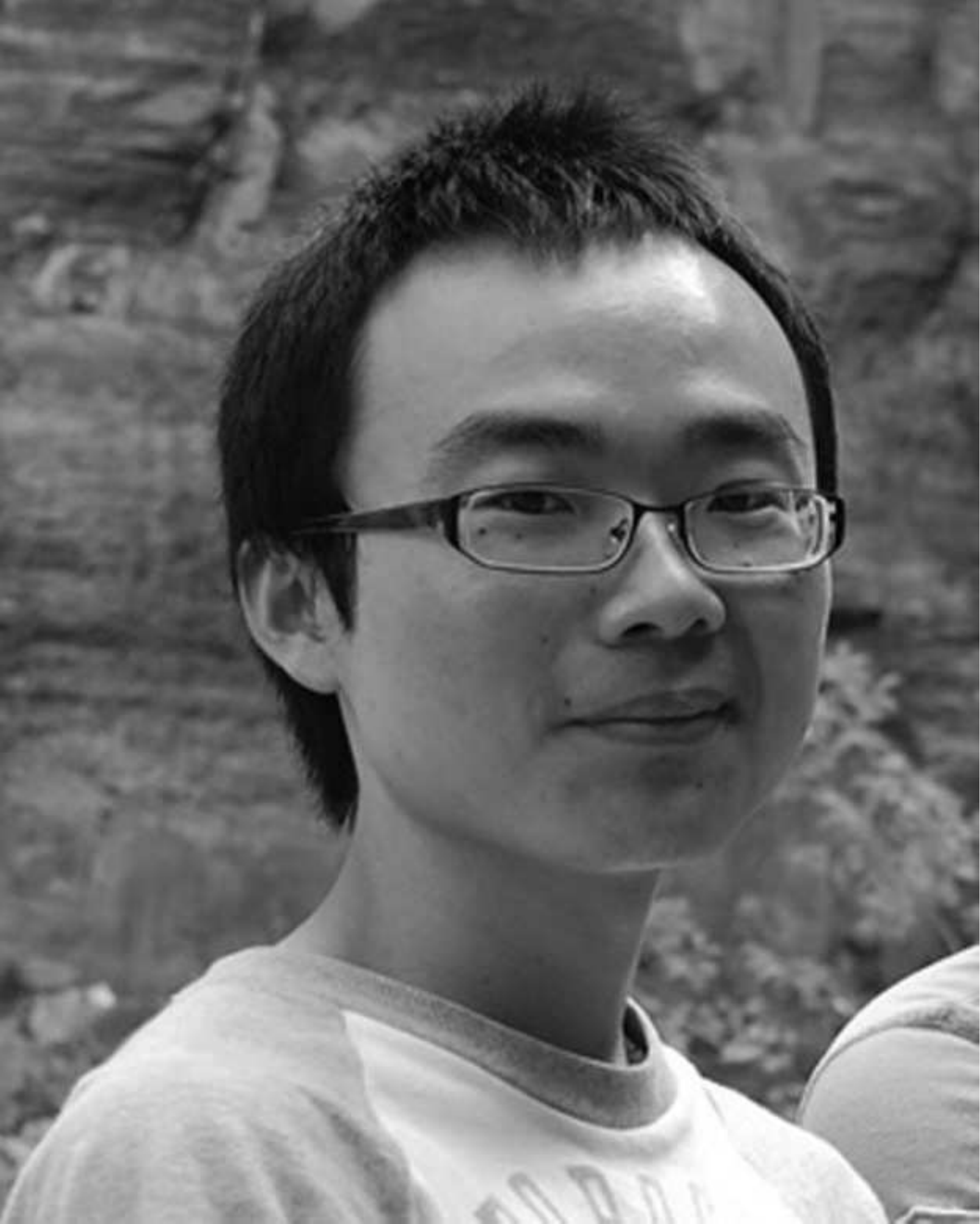}}]{Tao Wang}
(M'13) received the B.S. degree in mathematics from the University of Science and Technology of China, Anhui, China in 2006, and the Ph.D. degree in mathematics from the Pennsylvania State University, University Park, PA 16802, USA in 2011. Since 2012, he has been a visiting scientist at the School of Electrical and Computer Engineering, Cornell University, Ithaca, NY 14853, USA. His research interests include nonlinear systems and control, mathematical programming, geometric analysis, and applications in electrical engineering and management of invasive disasters.

\end{IEEEbiography}

\begin{IEEEbiography}[{\includegraphics[width=1in,height=1.25in,clip,keepaspectratio,angle=0]{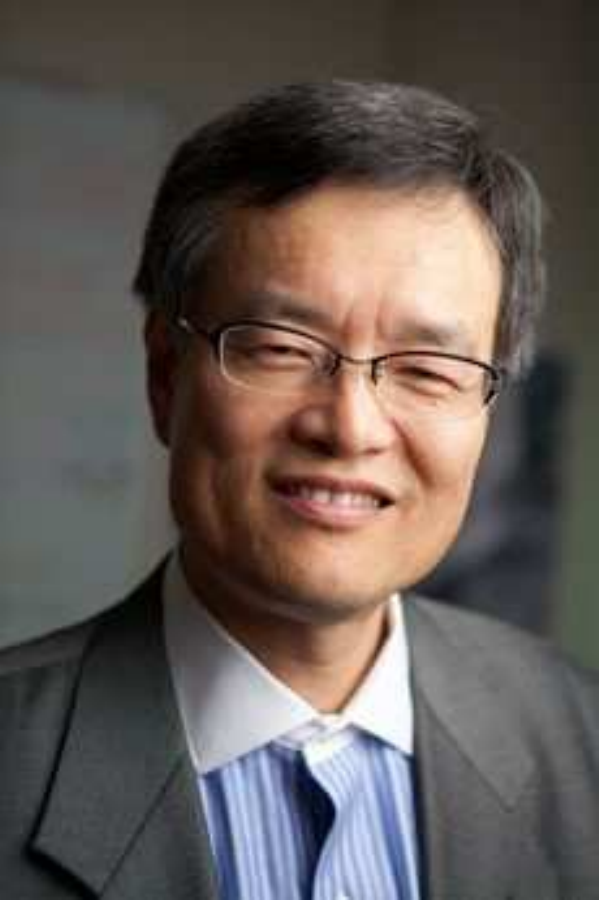}}]{Hsiao-Dong Chiang}
(M'87-SM'91-F'97) received the Ph.D. degree in EECS from the University of California at Berkeley in 1986. He is Professor of Electrical Engineering at Cornell University, Ithaca, New York. He and his research team have published more than 350 papers in refereed journals and conference proceedings. His current research interests include nonlinear system theory, nonlinear computation, nonlinear optimization and their practical applications. He was an associate editor IEEE Transactions on Circuits and Systems (1990-91, 1993-1995). He holds 17 U.S. and oversea patents and several consultant positions. He is Author of the book ``Direct Methods for Power System Stability Analysis: Theoretical Foundation, BCU Methodology and Applications'', John Wiley \& Sons, 2011 and of the book ``Stability region of nonlinear dynamical system: theory, optimal estimation and applications'', Cambridge Press, 2015.
\end{IEEEbiography}

\begin{IEEEbiography}[{\includegraphics[width=1in,height=1.25in,clip,keepaspectratio,angle=0]{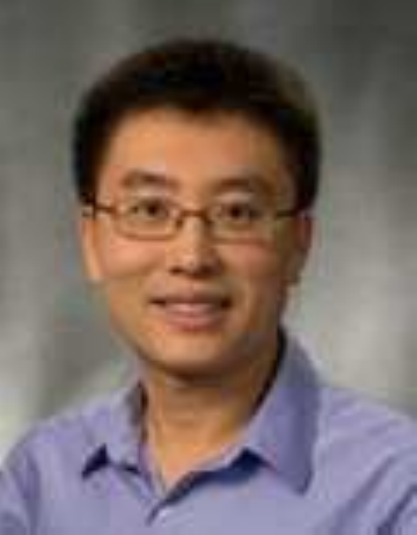}}]{Jianhui Wang}(M’07-SM’12 received the Ph.D. degree in electrical engineering from Illinois Institute of Technology, Chicago, IL, USA, in 2007. Presently, he is the Section Lead for Advanced Power Grid Modeling at the Energy Systems Division at Argonne National Laboratory, Argonne, IL, USA. Dr. Wang is the secretary of the IEEE Power \& Energy Society (PES) Power System Operations, Planning \& Economics Committee. He is an associate editor of Journal of Energy Engineering and an editorial board member of Applied Energy. He is also an affiliate professor at Auburn University and an adjunct professor at University of Notre Dame. He has held visiting positions in Europe, Australia and Hong Kong including a VELUX Visiting Professorship at the Technical University of Denmark (DTU). Dr. Wang is the Editor-in-Chief of the IEEE Transactions on Smart Grid and an IEEE PES Distinguished Lecturer. He is also the recipient of the IEEE PES Power System Operation Committee Prize Paper Award in 2015.
\end{IEEEbiography}

\begin{IEEEbiography}[{\includegraphics[width=1in,height=1.25in,clip,keepaspectratio,angle=0]{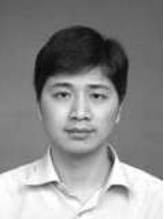}}]{Hui Liu}
(M'12) was born in Sichuan, China. He received the M. S. and Ph.D. degrees from the college of electrical engineering, Guangxi University, China, both in Electrical Engineering. He worked in Tsinghua University as a postdoctoral fellow from 2011 to 2013. He is an associate professor of school of electrical and information engineering, Jiangsu University, China. His research interests include power system control and electric vehicles.
\end{IEEEbiography}

\end{document}